\documentclass[aps,twocolumn,floatfix,longbibliography]{revtex4-1}
\usepackage{amsmath}
\usepackage{amsthm}
\usepackage{amsfonts}
\usepackage{amssymb}
\usepackage{subfigure}
\usepackage{appendix}
\usepackage{verbatim}
\usepackage{epsfig}
\usepackage{nameref}
\usepackage{algpseudocode}
\usepackage{newfloat}
\usepackage{float}
\usepackage{color}
                                                        
\newcommand{\vect}[1]{\boldsymbol{#1}}
\newcommand{\dif}{\mathrm{d}}

\usepackage{array}

\usepackage{newfloat}

\DeclareFloatingEnvironment[
    fileext=loa,
    listname=List of Algorithms,
    name=ALGORITHM,
    placement=tbhp,
]{algorithm}
\makeatletter
\let\OldStatex\Statex
\renewcommand{\Statex}[1][3]{%
  \setlength\@tempdima{\algorithmicindent}%
  \OldStatex\hskip\dimexpr#1\@tempdima\relax}
\makeatother
\algnewcommand{\LineComment}[1]{\Statex[0] \(\triangleright\)  #1 \hfill~}

\newcommand{\To}{\textbf{to~}}
\newcommand{\Is}{\textbf{is~}}
\newcommand{\False}{\textbf{False~}}
\newcommand{\LList}{\textsf{LiveList~}}
\newcommand{\OList}{\textsf{OutputList~}}
\newcommand{\db}{\textsf{minima.db~}}
\newcommand{\In}{\textbf{in~}}
\newcommand{\sens}{\texttt{MCLoop~}}

\begin{document}

\title{Superposition Enhanced Nested Sampling}

\author{Stefano \surname{Martiniani}$^{\dagger}$}
\email{sm958@cam.ac.uk}
\thanks{Corresponding author}
\author{Jacob D. \surname{Stevenson}}
\thanks{These authors have contributed equally to this work}
\author{David J. \surname{Wales}}
\author{Daan \surname{Frenkel}}

\affiliation{University Chemical Laboratories, University of Cambridge, Lensfield Road, Cambridge CB2 1EW, UK}

\date{\today}

\begin{abstract}
The theoretical analysis of many problems in physics, astronomy and applied mathematics requires an efficient numerical exploration of multimodal parameter spaces that exhibit broken ergodicity. Monte Carlo methods are widely used to deal with these classes of problems, but such simulations suffer from a ubiquitous sampling problem: the probability of sampling a particular state is proportional to its entropic weight. Devising an algorithm capable of
sampling efficiently the full phase space is a long-standing problem. Here we report a new hybrid method for the exploration of multimodal parameter spaces exhibiting broken ergodicity. Superposition enhanced nested sampling (SENS) combines the strengths of global optimization with the unbiased/athermal sampling of nested sampling, greatly enhancing its efficiency with no additional parameters. We report extensive tests of this new approach for atomic clusters that are known to have energy landscapes for which conventional sampling schemes suffer from broken ergodicity. We also introduce a novel parallelization algorithm for nested sampling.
\end{abstract}

\maketitle

\section{Introduction}
Computer simulations play an important role in the study of phase transitions
and critical phenomena. In particular, stochastic techniques such as Monte
Carlo (MC) methods have proved to be powerful tools~\cite{Landau2005}.
These methods rely on the ability of the Monte Carlo algorithm to sample the accessible volume in phase space.
There are, however, many situations where standard Monte Carlo simulations suffer from a lack of ergodicity. 
In that case, more sophisticated algorithms are needed to explore the volume in phase space that is, in principle, accessible.
Some such techniques are based on the efficient exploration of the underlying, multi-dimensional potential energy surface (PES)~\cite{Wales2003}. 
The PES, or energy landscape, can be viewed as a collection of basins separated by
barriers, where each basin corresponds to a particular local minimum-energy
configuration. The basin volumes define the entropic weight of the corresponding
local minima. The transition rate from one basin to another depends 
on the barrier height as well as the relative entropic weights (configurational 
space volumes)~\cite{Wales2003}. Many PES of interest exhibit frustration in 
the form of low-lying minima with different morphologies separated by high barriers. 
These structures may act as kinetic traps, when fixed-temperature sampling
methods such as molecular dynamics or Metropolis Monte Carlo sampling are used. 
There exist a wide range of extended or biased sampling techniques, both in Monte Carlo and in Molecular Dynamics,
that make it possible to speed up the sampling of landscapes with kinetic traps.
These techniques include Monte Carlo methods, such as umbrella sampling~\cite{Torrie1977,Berg1992}, density of states based methods, such as the 
Wang-Landau method~\cite{Wang2001} and replica exchange methods~\cite{Swendsen1986, Swendsen1987}, along with their Molecular Dynamics counterparts. Examples are the replica-exchange MD method~\cite{Sugita1999}, and the Meta-Dynamics method~\cite{Laio2002}. In cases where a biased distribution is generated, the original distribution can be reconstructed using reweighting techniques~\cite{Ferrenberg1988,Chodera2007} However, these approaches may perform poorly when dealing with high-dimensional spaces exhibiting broken ergodicity or, in other words, with highly multimodal (or multifunnel~\cite{Wales2003,Wales2013,Doye1999,Doye1998}) parameter spaces~\cite{Poulain2006,MANDELSHTAM,SHARAPOV,Sharapov2007}.

In recent years a Bayesian method known as nested sampling~\cite{Skilling2006}
has emerged as a possible alternative to extended or biased sampling methods. The nested-sampling approach has found widespread application in astrophysics~\cite{Mukherjee2006a,Feroz2008} and cosmology~\cite{Shaw2007a,Feroz2009}, and has drawn the attention of computational and statistical physicists~\cite{murray2006,Partay2010,Nielsen2013,Burkoff2012,Partay2012,Brewer2010}. Furthermore, the method has been recently adopted for Bayesian model comparison in systems biology~\cite{Aitken2013,Dybowski2013,Pullen2014}. Nested sampling explores phase space in an unbiased way, and allows one to determine statistically the density of states associated with shrinking fractions of phase space. This objective is achieved by placing a constraint on the potential energy (for instance), which decreases at each nested sampling iteration. Like Wang-Landau sampling, the method is athermal and produces the density of states and the partition function (Bayesian evidence) as its primary product. However, nested
sampling does not require binning of the energy for systems with continuous
potentials. The self-adapting steps in energy (but constant in $\log$ phase
space volume) is attractive because the approach does not require prior knowledge of possible phase
transitions. For example, the step size adjusts automatically as the
phase space volume shrinks near a first order phase transition~\cite{Skilling2006,Partay2010}. 

An important drawback of nested sampling is that when the decreasing
energy constraint forbids a transition to an unexplored basin, that basin
cannot be visited and ergodicity is broken. Hence, while nested sampling
certainly is conceptually interesting, its performance is often no better than that of conventional extended sampling methods in
dealing with systems exhibiting broken ergodicity~\cite{Partay2010}. 
In the present work we introduce a novel hybrid methodology for 
the exploration and thermodynamic analysis of such systems. 

Superposition enhanced nested sampling (SENS) combines the strengths of
unbiased global optimization techniques~\cite{Wales2003} with those of nested
sampling. Global optimization techniques such as basin-hopping~\cite{Li1987,Wales1997,C3CP44332A} are designed to find the lowest energy configuration of
a PES. They are not constrained to sample according to any distribution, so
they are free to use `quick and dirty' techniques while searching
for the global minimum. For example, they can take Monte Carlo steps that
do not satisfy detailed balance, and make use of minimisation algorithms such as
L-BFGS and conjugate gradient. Such operational freedom makes global
optimization algorithms much more efficient than generalised ensemble methods
at locating the lowest energy
minima~\cite{mcginty71,burton72,hoare79,stillingerw84,Strodel2008}. 
Collections of the lowest energy minima configurations thus obtained can then be used in
the context of the superposition approach (SA)~\cite{Wales2003,stillingerw84,Strodel2008,Wales1993,Doye1995,Doye1995a,CalvoDW01,georgescu:144106}
to compute the thermodynamic properties of the system. However, doing so
accurately at high temperatures using the SA alone often requires a
prohibitively large number of minima.

In the present contribution we show how knowledge of the lowest energy 
minima and their statistical weights, calculated
using the harmonic superposition approximation (HSA), 
can be exploited to enhance the problematic low energy
behaviour of nested sampling, thus making it likely that none of the important
minima and associated regions are missed. Although we discuss SENS in the context of
energy landscapes, the method is completely general and can be applied to
any multi-modal parameter space whose minima (maxima in likelihood) can be
identified by global optimization algorithms.

\section*{Nested Sampling}

Nested sampling~\cite{Skilling2006} provides an elegant solution to the problem of
evaluating the density of states, and hence the partition function, for arbitrary
systems. A likelihood value is assigned to each possible configuration. For our purposes
the likelihood is the Boltzmann factor $\exp(-E / k_B T)$, but it could 
be some other measure. Typically, there are large numbers of configurations with a low likelihood. In addition, there may be a small number of configurations with high likelihood.

The aim of nested sampling is to sample configuration space uniformly, but with the energy constrained to lie below a maximum value, $E^{max}$, that decreases iteratively throughout the calculation. The rate of decrease is maintained self-consistently, such that the phase space volume with energy less than $E^{max}$ decreases by a constant factor in each iteration.
 
The nested sampling algorithm starts by generating $K$ configurations of the
system completely at random, distributed uniformly, in configuration space. 
The energy, $E_\mathcal{R}$, of each of these configurations is computed and added to a sorted list, where $\mathcal{R}$ is the associated index in the sorted list. For each of these replicas we define the configurational phase space volume, $\Omega_{E\leq E_\mathcal{R}}$, containing all configurations with $E \leq E_\mathcal{R}$. The key insight of nested sampling is that the volumes, $\Omega_{E \leq E_\mathcal{R}}$, normalised by the total phase space volume, are distributed according to the Beta distribution, $\text{Beta}(K-\mathcal{R}+1,\mathcal{R})$ \footnote{The normalised fraction of configuration space $X_i= \Omega_{E\leq E_{\mathcal{R}}}/\Omega_{tot} \in (0,1)$. Since Nested Sampling assumes uniform sampling in phase space, estimating $X_i$ becomes analogous to estimating the $\mathcal{R}$-th largest position out of $K$ points sampled at random on a unit segment. The point with $\mathcal{R}$-th largest position will be distributed on the unit line as $\text{Beta}(K-\mathcal{R}+1,\mathcal{R})$.}. This distribution has expectation value and variance
\begin{equation}
\label{eq:gen_expectation}
\mu_{\mathcal{R}} = 1-\frac{\mathcal{R}}{K+1}~~\text{and}~~\sigma_{\mathcal{R}}^2 = \frac{\mathcal{R}(K-\mathcal{R}+1)}{(K+2)(K+1)^2}.
\end{equation}
The above formalism assumes that the total phase space volume $\Omega_{tot}$ is finite, 
but this condition can generally be satisfied with negligible error, 
for example by placing the system in a large box.

At the $i$-th nested sampling iteration the replica (out of $K$
replicas) with highest energy $E^{max}_i$ is discarded and replaced by a new configuration 
sampled uniformly under the constraint $E \leq E^{max}_i$. 
The maximum energy $E^{max}_i$ is stored for later analysis.
Again, the volume of configuration space with energy less than the $\mathcal{R}$-th
largest energy, $\Omega_{E \leq E_\mathcal{R}}$,
this time normalized by $\Omega_{E \leq E^{max}_i}$, is distributed according to the Beta 
distribution with mean and variance given by Eq.~(\ref{eq:gen_expectation}).
During the nested sampling iteration
the volume of phase space with energy below $E^{max}$ contracts, on average, by $\mu_{1} = K/(K+1)$. After $N$ nested sampling iterations, the algorithm produces a list of the form ${\{E^{max}_1,E^{max}_2,\dots,E^{max}_N\}}$. We can associate a fraction of configuration space 
$X_i = \Omega_{E \leq E^{max}_i} / \Omega_{tot} = \mu_1^i$, with each $E^{max}_i$.
The density of states, or the (normalized) volume of phase space with energy between $E^{max}_{i+1}$ and $E^{max}_{i}$ is simply
\begin{equation}
g_i(E) = X_i - X_{i+1} = \mu_1^i - \mu_1^{i+1} = \frac{1}{K+1} \left( \frac{K}{K+1}\right)^i.
\end{equation}
Thermodynamic quantities of interest, such as the mean energy, entropy, free energy, and 
heat capacity, can easily be computed from the density of states at arbitrary temperature.

To generate configurations uniformly in space we use
the strategy suggested by Skilling~\cite{Skilling2006}: after removing the
configuration with highest energy one of the remaining $K-1$ replicas (chosen randomly)
is duplicated.
The new configuration is then evolved through a Markov chain Monte Carlo (MCMC) walk 
sufficiently long to decorrelate the system from its initial state. 
This Monte Carlo walk is equivalent to a normal Monte Carlo simulation 
at infinite temperature. The coordinates are randomly
perturbed, and the new configuration is accepted subject only to the condition
that the energy remains below $E^{max}$. For most systems of interest the vast 
majority of the computational effort will be spent generating new configurations.

\subsection*{Parallelization}
\label{sec:parallelisation}
Nested sampling can be formulated to run in parallel on an
arbitrary number of processors.
We present a pseudocode description of our parallel implementation in
Algorithm~\ref{alg:nested_sampling}. Since this scheme also constitutes the basic
framework for SENS we define the MCMC loop in the most
general way at line $9$ of Algorithm~\ref{alg:nested_sampling}. For the
purpose of discussing the algorithm in its simplest form here we
will consider Algorithm~\ref{alg:simplest}.

\begin{algorithm}[ht]
\begin{algorithmic}
\For{$l=0$ to $N$-steps}
\State generate trial configuration (e.g. by random 
\Statex[1] perturbation);
\State if $E_{trial} \leq E_{max}$: accept trial configuration;
\EndFor
\end{algorithmic}
\caption{Nested sampling \texttt{MCLoop}}
\label{alg:simplest}
\end{algorithm}

At each nested sampling iteration, instead of removing only the replica with the 
highest energy, we remove the $\mathcal{P}$ replicas with highest energy,
where $\mathcal{P}$ is the number of processors available. 
The rate of phase space contraction now is given by $\mu_{\mathcal{P}}$, 
leading to much faster phase space contraction and shorter calculations in 
terms of wall-clock time. This parallelization procedure was
first described in reference~\cite{Burkoff2012}. Our improvement is that we do not discard
the $\mathcal{P}-1$ replicas with highest energy but we store them for later analysis. 
Phase space contraction between iterations is still constant, but now the post-analysis is
slightly more complicated.
The fraction of configuration space associated with the
$n$-th recorded energy is
\begin{equation}
X_n = \prod_{i=0}^{n} \frac{K-i\%\mathcal{P}}{K+1-i\%\mathcal{P}},
\end{equation}
where ``$\%$'' is the mod operator. 
This method follows
the same stepping routine as the existing parallelization algorithm. However it
produces $\mathcal{P}$ times as many points, hence providing a more detailed
picture of the potential energy surface and a much more fine-grained 
binning of the density of states.

\begin{algorithm}[ht]
\begin{algorithmic}[1]
\LineComment{initialisation}

\State generate $K$ random configurations;

\State store their coordinates and energy in \LList;

\LineComment{main loop}

\While{termination condition \Is~\False}

\State remove the $\mathcal{P}$ replicas $\{ {\bf R}_{m}^{(1)},\dots , {\bf R}_{m}^{(\mathcal{P})}\} \equiv \{ {\bf R}_{m} \}$ 
\Statex[1] with highest energy $\{ E_{m}^{(1)} > \dots > E_{m}^{(\mathcal{P})} \} \equiv \{ E_{m} \}$ 
\Statex[1] from \LList;

\State append $\{ E_{m} \}$ to \OList; 

\State set $E_{max} = E_{m}^{(\mathcal{P})}$;

\State select $\mathcal{P}$ replicas $\{ {\bf R}_{s}^{(1)},\dots,{\bf R}_{s}^{(\mathcal{P})} \} \equiv \{{\bf R}_s\} $ from
\Statex[1] \LList at random;

\State add a copy of $\{{\bf R}_s\}$ to \LList;

\State \texttt{MCLoop}\{$\{{\bf R}_s\}$,~$E_{max}$,~\db\}

\EndWhile

\end{algorithmic}
\caption{Parallel nested sampling}
\label{alg:nested_sampling}
\end{algorithm}


\section*{SENS - the concept}
Global optimization is a common numerical problem and global optimization algorithms have been developed
in many areas of science~\cite{Hartke2011,Wales1999,Wales2003}. Knowledge of the local minima alone, however, is not sufficient to infer all the observable properties of interest 
from the energy landscape (or in general any parameter space). The harmonic
superposition approximation (HSA)~\cite{Stillinger1984} (for more details, see e.g.~\cite{Wales2003}) allows one to compute the density of states and the partition function, solely based on the knowledge of the individual local
minima and the local curvatures (normal mode frequencies) 
of the potential energy landscape, via the Hessian matrix. In the
HSA each local minimum corresponds to a harmonic basin and observable
properties are expressed as a sum over individual contributions of the minima.

The HSA has been shown to be very effective for several systems~\cite{Sharapov2007,Somani2013,Strodel2008}
but the accuracy depends on how well the potential energy of the basins
can be approximated as harmonic, and how many minima are thermodynamically
important. While the HSA captures landscape anharmonicity, arising from the distribution of local minima, it does not include well anharmonicity, arising from the shape of the well. Therefore, the HSA becomes an increasingly good approximation at lower energies where well anharmonicity is less important. The total number of minima increases exponentially with system size~\cite{Stillinger1984,WalesD03}, but it is impossible to tell a-priori how many of those are important.
For example, LJ$_{31}$, a cluster of $31$ isotropic particles interacting
through a Lennard-Jones potential~\cite{Lennard-Jones1931}, 
has about $3 \times 10^{15}$ distinct minima~\cite{Wales2013}, but only a few dozen are required to reproduce 
the low temperature thermodynamic behaviour.

The global resolution of nested sampling depends on the number of replicas, $K$,
used in the simulation, which is generally limited by the
available computation time (the larger $K$, the slower the descent in energy). A more
serious problem for nested sampling is that if the barrier to enter an
unexplored funnel or superbasin is higher than the energy constraint $E_{max}$, that
region of the PES will never be explored if it is not already populated.
For example, in a crystallisation transition, at high energy the
statistical weight of the liquid phase will be overwhelming and there will be
no replicas in the region corresponding to the solid phase. However, as the
energy constraint decreases (hence the temperature) the relative statistical weight
associated with the solid phase increases. If we could sample phase space
uniformly then at low energy we would observe a phase transition corresponding to
crystallisation, but we must resort to a MCMC walk 
to explore phase space. If the entrance to the superbasin
corresponding to the crystal has been locked out by $E_{max}$ a Markov chain
will not be able to find it, thus missing the transition.

Here we propose a new method that combines complementary
techniques: nested sampling can sample efficiently the high energy regions of
phase space, while at low energy a database of minima obtained by global optimization
is used to augment the survey. While nested
sampling assigns the correct statistical weight to each basin, global
optimization makes it likely that no important minima are missed. This philosophy is also used
in other methods combining replica exchange Monte Carlo with global optimization algorithms to treat broken ergodicity~\cite{Zhou1997,Ioan2001,MANDELSHTAM,Wales2013}.

\section*{SENS - the algorithm}

Employing knowledge of low-lying minima fits naturally within the
framework of nested sampling. We present here both an
exact and an approximate implementation of the SENS algorithm. Exact SENS is
fully unbiased and requires no additional parameters than those needed in
nested sampling. Approximate SENS, on the other hand, is formally biased and
requires additional parameters. The reason for presenting both methods is 
that, in some cases, the latter approach can be
considerably simpler to implement than the former, while generally producing
equally good, or better, results. SENS is based on the original nested sampling algorithm
presented in Algorithm~\ref{alg:nested_sampling}. The novelty of our method
resides in the augmented sampling of the parameter space obtained by coupling
the MCMC to the HSA. SENS can therefore be implemented by changing the function
\texttt{MCLoop}$\left( \{{\bf R}_s\},~E_{max},~\textsf{minima.db} \right)$ of
Algorithm~\ref{alg:nested_sampling}. A full outline of the SENS algorithm can
be found in Algorithm~\ref{alg:sens} of the Supplementary Information. To run SENS, a
database of the lowest energy minima must be pre-computed.\\
In this work we adopt basin-hopping~\cite{Li1987,Wales1997,C3CP44332A} as the global optimization algorithm of choice. Basin-hopping associates any given point of the PES to a minimum obtained by energy minimisation, thus transforming the PES into a set of catchment basins. This basin transformation is combined with a sampling scheme to search for the global minimum. At each step the coordinates of the current minimum configuration are perturbed to hop out the basin and minimised again to find a new minimum. Each step between two minima configurations is accepted with probability 
\begin{equation*}
P(\vect{x}_{old}\rightarrow \vect{x}_{new})=\text{min}\left[1,\exp(-\beta(E_{new}-E_{old})) \right].
\end{equation*}
If the move is rejected, the coordinates are reset to those of the current local minimum. Since perturbations should be large enough to step out of the catchment basin, the step-size is typically much larger than for thermodynamic sampling. Furthermore, since detailed balance need not hold, the step-size can be dynamically adjusted to improve sampling. Basin-hopping has been successfully applied to a wide range of atomic, molecular, soft and condensed matter systems~\cite{Mochizuki2014,Somani2013,Olesen2013,Forman2013}. 
\subsection*{Exact SENS}
An unbiased version of SENS can be implemented by means of Hamiltonian replica
exchange Monte Carlo moves~\cite{Bunker2000,Fukunishi2002}: in addition to normal MC steps, we introduce a new Monte Carlo step in which a minimum is sampled from the
database according to its HSA configurational entropic weight:
\begin{equation}
w_c^{(b)}(E) = \frac{\Omega_c^{(b)}(E)}{\Omega_c(E)}.
\label{eq:entropic_weight}
\end{equation}
We define the configurational volume of basin $b$
\begin{equation}
\Omega_c^{(b)}(E) \propto \frac{n_b (E - V^{(b)})^{\frac{\kappa}{2}}}{\prod_{\alpha=1}^{\kappa} \nu_{\alpha}^{(b)}},
\label{eq_basin_volume}
\end{equation}
and the total configurational volume
\begin{equation}
\Omega_c(E) \propto \sum_b \frac{n_b (E - V^{(b)})^{\frac{\kappa}{2}}}{\prod_{\alpha=1}^{\kappa} \nu_{\alpha}^{(b)}},
\label{eq_total_volume}
\end{equation}
where $V^{(b)}$ is the potential energy of the minimum corresponding to basin $b$, 
$\nu_{\alpha}^{(b)}$ are the normal mode vibrational frequencies defined by the Hessian matrix, $\kappa$ is the number of vibrational degrees of freedom, and $n_b$
is the degeneracy of the basin (for Lennard-Jones clusters this is the number of
distinct non-superimposable permutation-inversion isomers for minimum $b$)~\cite{Wales2003}. 
Here we have left out all the constant factors that cancel out as well as overall rotations. 
Once a minimum is selected, a configuration with $E \le E_{max}$ is then 
chosen uniformly from within its basin of attraction. This approach corresponds to selecting a point uniformly from a multidimensional harmonic well. Such a 
configuration can be generated analytically, see the Supplementary Information for details. Unlike Ref.~\cite{MANDELSHTAM}, in our approach we sample from the uniform distribution of configurations below energy $E_{max}$, rather than from the corresponding canonical distribution. 

Thus, we obtain a configuration ${\bf R}_{sys}$ sampled according to the true
 Hamiltonian, $\mathbb{H}_{sys}$, and a trial configuration, ${\bf R}_{har}$,
sampled according to the HSA-Hamiltonian, $\mathbb{H}_{har}$. The energies of the two
configurations are then computed with the other Hamiltonian. If 
\begin{equation}
\mathbb{H}_{har}({\bf R}_{sys}) \leq E_{max}~\text{and}~\mathbb{H}_{sys}({\bf R}_{har}) \leq E_{max},
\label{eq:detailed_balance}
\end{equation}
then the true distribution and the HSA distributions overlap, the swap is accepted, and ${\bf R}_{sys}$ becomes ${\bf R}_{har}$, otherwise it is rejected. This procedure guarantees that detailed balance is satisfied, for further discussion refer to the Supplementary Information. In practice only the lowest energy minima will successfully swap, since the HSA can only be reasonably accurate around these basins. It is, however, at low energy that such swaps are needed the most due to the hard energy constraint used by nested sampling. Note that swaps are complemented by regular MCMC walks to allow for the exploration of the full configuration space. In SENS the replicas are allowed to ``tunnel'' between basins, thus improving the sampling. A more detailed description, along with a pseudo-code implementation of \texttt{MCLoop} specific to Lennard-Jones clusters is provided in the Supplementary Information.

\subsection*{Approximate SENS}
\label{sec:approx_sens}
The implementation of approximate SENS is somewhat simpler, but comes at the
cost of at least one extra parameter. The basic idea of approximate SENS is that
the sampling of configuration space can be augmented by starting a MCMC walk from
a local minimum configuration, sampled from the database according to its entropic
weight Eq.~(\ref{eq:entropic_weight}), with some user defined frequency. This
frequency is intrinsically defined in exact SENS by the relative overlap of the
HSA and the true density of states. To implement approximate SENS
we only need a database of minima and their relative weights computed 
according to Eq.~(\ref{eq:entropic_weight}). 
Before each MCMC step a random number is drawn. If this number is
less than some user defined probability, $P_{DS}$, then a minimum
is selected from the database according to the HSA weights and the MCMC walk starts 
from this minimum configuration. A pseudo-code implementation of \texttt{MCLoop} 
for approximate SENS is provided in the Supplementary Information. 

There are two main sources of bias in the approximate SENS. The first one is
due to the limited number of minima from which we sample, since we cannot
include the large number of high energy minima. 
The second source of error is due to the poor quality of the
HSA approximation far from the minimum, hence the entropic weights for the minima are
not accurate at high energy. The most obvious way of reducing these
biases is to use long MCMC walks. In fact, if we sample from the wrong
basin a long MCMC walk will allow the system to escape and explore regions of phase
space with greater entropic weight. However, very long MCMC walks are
computationally expensive, and if short runs are required we need to
sample from the database of minima carefully. 
If we start sampling from the database of minima at high energy
we will possibly introduce a bias due to over-weighting of the low energy
regions of configuration space. 
To avoid this problem we suppress sampling from the database 
until we are sure the HSA is likely to describe the potential energy landscape accurately.
We use a simple function (of the Fermi type) that
delays the onset of sampling from the database of minima and limits its maximum
frequency 
\begin{equation}
\label{eq:onset_function}
f_{onset} = \frac{f_{max}}{1 + e^{(E_{min}^{(\mathcal{R})}-E_{on})/\Delta E}},
\end{equation}
where $E_{on}$ is some onset energy and $E_{min}^{(\mathcal{R})}$ is 
the energy of the replica with lowest energy. $E_{on}$ 
could be chosen as $E_{max}^{(\textsf{minima.db})}$, the energy of the highest known
minimum (stored in the database), or as the largest energy at which the HSA describes the system accurately. $f_{max}$ and $\Delta E$
are user-defined parameters that determine the total probability of
sampling after the onset and the width of the onset region, respectively. 
For small sampling probabilities, $P_{DS} \ll 1$, the optimal frequency of sampling from the database, should scale as $1/K$; a theoretical justification is derived in the
Supplementary Information. Hence, for $P_{DS} \ll 1$, we can make the probability of sampling from the database independent of the number of replicas, replacing $f_{max}$ with $f_{max}/K$.

We identify two possible strategies for the application of
approximate SENS. One is to start sampling from a large database 
early in the simulation when ${E_{on}=E_{max}^{(\textsf{minima.db})}}$, 
with a small $P_{DS}$, hence we choose $f_{max} \ll 1$. 
This procedure allows nested sampling
to do most of the work, but ensures that no important basins will be missed. 
Alternatively, sampling from the database can be delayed until all the high temperature 
transitions have occurred, at which point we start
sampling more extensively from the database, hence $f_{max} \gtrsim 1/2$.
Note that the database can be considerably smaller in this case. 
The first strategy is a slight enhancement to nested sampling, while the
latter strategy interpolates between nested sampling and the HSA in a similar
spirit to the basin-sampling method~\cite{Wales2013}. Importantly, even if we sample from
the database of minima, we use the MCMC walk to explore more the anharmonic regions of a 
basin, allowing us to go beyond the harmonic approximation.

\section*{Results}

We test SENS by calculating the thermodynamic properties
of Lennard-Jones clusters exhibiting broken ergodicity. Lennard-Jones (LJ) clusters
are systems of particles that interact via the Lennard-Jones potential\cite{Lennard-Jones1931}
\begin{equation}
E = 4 \epsilon \sum_{i<j}\left [ \left(\frac{\sigma}{r_{ij}} \right)^{12} - \left(\frac{\sigma}{r_{ij}} \right)^6 \right ],
\end{equation}
where $\epsilon$ is the pair well depth, $\sigma$ is the separation at which
$E=0$, and $2^{1/6}\sigma$ is the equilibrium pair separation. LJ clusters have
served as benchmarks for many global optimization techniques and thermodynamics sampling~\cite{Wales1997,Wales2003,Wales2013,Partay2010,Sharapov2007,Poulain2006}.

The majority of putative ground states for LJ clusters are based 
on icosahedral packings~\cite{Doye1999}. For some \emph{magic number} LJ
clusters complete Mackay icosahedra are possible, for examples $N=13,55$.
Complete icosahedral structures are considerably more stable than 
neighbouring sizes and their landscape is funneled towards the global minimum~\cite{Doye1999}.
There are, however, other sizes for which the global minimum is not
icosahedral. Examples are LJ$_{38}$, whose ground state is an fcc-truncated
octahedron~\cite{Doye1999}, and LJ$_{75}$ whose global minimum is a
Marks decahedron~\cite{Doye1999}. Due to the overwhelming number (entropic
weight) of structures based on incomplete icosahedra at high energy, the energy
landscapes of LJ clusters with nonicosahedral global minima exhibit broken ergodicity. 
Calculating accurate thermodynamic properties
for these systems has proved to be a real challenge for all conventional
techniques~\cite{Sharapov2007,Poulain2006,Wales2003,Partay2010} and hybrid or more
complicated schemes~\cite{Wales2013,Sharapov2007,Poulain2006,Doye1998} are
necessary. LJ clusters with broken ergodicity therefore provide excellent benchmarks to
test the performance of new sampling techniques.

\subsection*{LJ$_{\bf 31}$}

\begin{figure}[t]
\centering
\includegraphics[width=\linewidth]{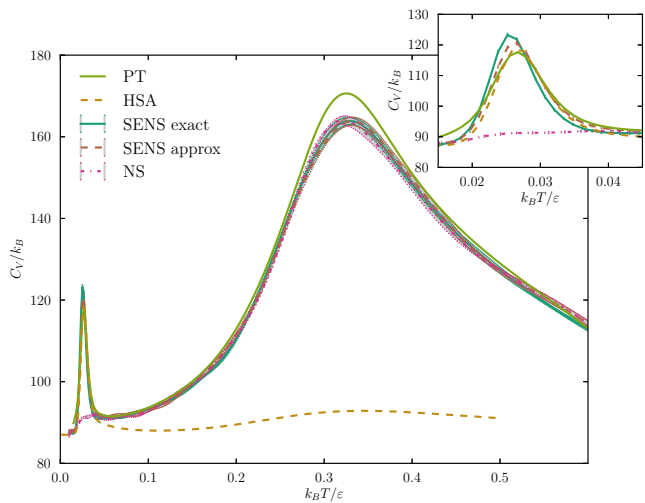}
\caption{Heat capacity curves for LJ$_{31}$. PT and HSA correspond to parallel tempering and the harmonic superposition approximation, respectively. All SENS calculations were performed using $K=20000$ replicas.}
\label{fig:cv_lj31}
\end{figure}

\begin{figure}[t]
\centering
\includegraphics[width=\linewidth]{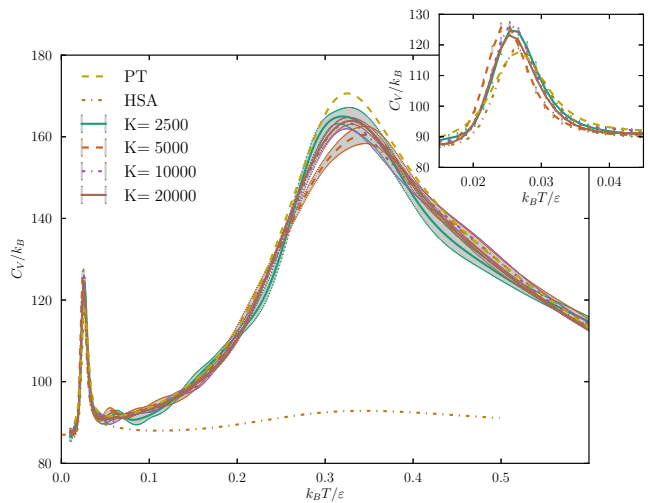}
\caption{Comparison of heat capacity curves for LJ$_{31}$ obtained by exact SENS using different numbers of replicas. The PT and HSA curves were obtained by parallel tempering and the harmonic superposition approximation, respectively.}
\label{fig:cv_lj31_comp_exact}
\end{figure}

\begin{table}[t]
\centering
\begin{tabular}{lccc}
\firsthline
\hline
\multicolumn{4}{c}{LJ$_{31}$} \\
\hline
Method      & $K$         & $N$        & $N_{\text{E}}^{(total)}$ \\
\hline
\hline
PT          &              &               & $1.9 \times 10^{11}$ \\
NS ref.\cite{Partay2010} & $280000$    &  & $3.4 \times 10^{12}$ \\
NS          & $20000$     & $10000$      & $1 \times 10^{11}$               \\
SENS approx & $20000$     & $10000$      & $1 \times 10^{11}$ \\
SENS exact  & $20000$     & $10000$      & $1 \times 10^{11}$  \\
SENS exact  & $10000$     & $10000$      & $5.2 \times 10^{10}$  \\
SENS exact  & $5000$     & $10000$      & $2.6 \times 10^{10}$  \\
SENS exact  & $2500$     & $10000$      & $1.3 \times 10^{10}$  \\
\lasthline
\end{tabular}
\caption{Comparison of methods used to obtain the LJ$_{31}$ heat capacity curves shown in Figs.~\ref{fig:cv_lj31}~and~\ref{fig:cv_lj31_comp_exact}. $N_{\text{E}}^{(total)}$ indicates the total number of energy evaluations (summed over all processors). PT was performed using $24$ replicas spread geometrically through the temperature range $0.0125$ to $0.6$. Note that approximate SENS can perform as well as exact SENS when fewer replicas are used, in the interest of brevity we do not include these results as the LJ$_{75}$ calculations illustrate clearly the capabilities of the method.}
\label{table:lj31}
\end{table}

LJ$_{31}$ is the smallest Lennard-Jones cluster exhibiting broken ergodicity and a low temperature solid-solid phase-like transition from Mackay to anti-Mackay surface structures~\cite{Doye1999}. Convergence of the heat capacity curve for LJ$_{31}$ by parallel tempering (PT) with $24$ geometrically distributed temperatures in the range $0.0125$ to $0.6$ required $N_E^{total} = 1.9 \times 10^{11}$ energy evaluations to converge (curve shown in Fig.~\ref{fig:cv_lj31}). Partay et al.~\cite{Partay2010} report that $K=288000$ replicas and $N_E^{\text{total}} = 3.4 \times 10^{12}$ energy evaluations were needed to converge the heat capacity curve of LJ$_{31}$ by nested sampling (NS) using a low particle density of $2.31 \times 10^{-3}\sigma^{-3}$ (100 fold less dense than our system). Fig.~\ref{fig:cv_lj31} compares the heat capacity curves obtained by PT, HSA (computed using $\gtrsim 80000$ minima), NS and SENS for LJ$_{31}$. The SENS and NS results correspond to $K=20000$ replicas, $N=10000$ steps for each MCMC walk, and $\mathcal{P}=16$ cores. The database of minima used for SENS contained the lowest $183$ minima, although for SENS exact we observe that only seven minima contribute to the swaps; see Table~IV of Supplementary Information for the swap statistics. From Fig.~\ref{fig:cv_lj31} we see that both exact SENS and approximate SENS are well converged and agree with the PT curve over the whole temperature range, and with the HSA at low temperature. We note that $K=20000$ replicas are not nearly enough for NS to converge, and the low temperature peak is in fact completely absent. Using this number of replicas SENS requires half the total number of energy evaluations of PT and one order of magnitude less than NS, see Table~\ref{table:lj31}. The swap operations do not constitute a noticeable overhead and the reduction in the total number of energy evaluations corresponds to an equivalent reduction in wall-clock time.

In Fig.~\ref{fig:cv_lj31_comp_exact} we show a comparison of PT, HSA and exact SENS for a range of replica numbers $2500 \leq K \leq 20000$; see Table~\ref{table:lj31} for comparison. We observe that the high temperature peak practically converges for $K=10000$ and it resembles the features of the converged curve quite well even for smaller numbers of replicas. The low temperature peak instead converges very quickly, for as few as $K=2500$ replicas, representing an improvement in performance of 20 times over PT. We note that one of the great strengths of SENS is that even when a small number of replicas are used and run times are very short, although the curves may not be completely converged, the physical picture produced by the method is always correct because all the important basins are visited. On the other hand, rapid convergence of the heat capacity curves, requires the HSA to be a good representation for the system. LJ$_{38}$ is an example for which this condition does not hold as well, see the Supplementary Information for further details.

\subsection*{LJ$_{\bf 75}$}

\begin{figure}[t]
\centering
\includegraphics[width=\linewidth]{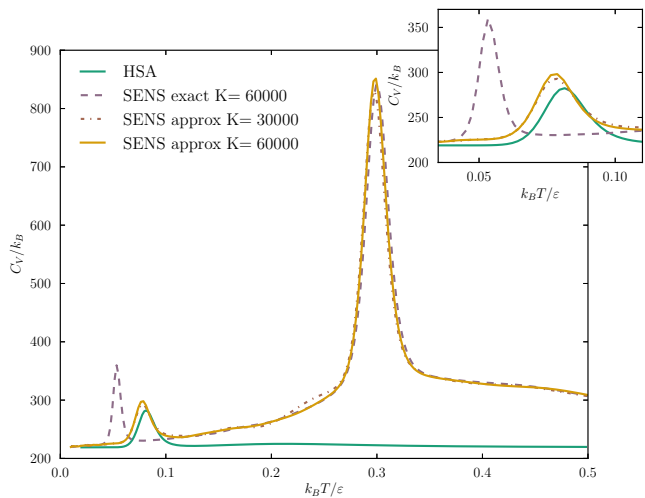}
\caption{Heat capacity curves for LJ$_{75}$. The PT and HSA results were obtained by parallel tempering and the harmonic superposition approximation, respectively. Exact SENS calculations were performed using $K=60000$ replicas, while results for approximate SENS calculations are shown for both $K=30000$ and $K=60000$ replicas.}
\label{fig:cv_lj75}
\end{figure}

LJ$_{75}$ is a particularly clear example of a double-funneled energy landscape~\cite{Doye1999} with $O(10^{25})$ distinct local minima~\cite{Wales2013}. The decahedral global minimum is separated by a very large potential energy barrier from the lowest icosahedral minimum. Sharapov and Mandelshtam~\cite{MANDELSHTAM} showed that $O(10^{12})$ (total) energy evaluations of adaptive parallel tempering are not enough to converge the heat capacity peak corresponding to the solid-solid phase-like transition in LJ$_{75}$~\cite{MANDELSHTAM}. Furthermore, the rate of convergence slows down dramatically (it practically stops) after $O(10^{11})$ (total) energy evaluations and coupling of PT to the HSA is necessary to obtain convergence of the low temperature peak~\cite{MANDELSHTAM}. Fig.~\ref{fig:cv_lj75} compares the heat capacity curves obtained by HSA (computed using $758$ minima) and SENS for LJ$_{75}$. SENS was carried out using $K=30000$ or $K=60000$ replicas and $N=10000$ steps for each MCMC walk on $\mathcal{P}=16$ processors. The database of minima for SENS contained the lowest $758$ minima. Approximate SENS started sampling from the database at $E_{on}=-369~\epsilon$, while for exact SENS only 10 of the minima contributed to the swaps; see Table~VI of the Supplementary Information for swap statistics. Unlike adaptive PT~\cite{MANDELSHTAM}, approximate SENS converges in $O(10^{11})$ energy evaluations (Table~\ref{table:lj75}), but exact SENS fails to converge the low temperature peak for the same number of replicas. As for LJ$_{38}$, exact SENS does not converge quickly due to the lower accuracy of the HSA, as inferred from the extremely low swap acceptance (Table~VI of Supplementary Information). On the other hand approximate SENS performs considerably better than for LJ$_{38}$ because the melting transition is well separated from the solid-solid transition, thus allowing sampling from the database relatively early on in the simulation (right after melting) without affecting the melting transition.

\begin{table}[t]
\centering
\begin{tabular}{lccc}
\firsthline
\hline
\multicolumn{4}{c}{LJ$_{75}$} \\
\hline
Method      & $K$         & $N$        & $N_{\text{E}}^{(total)}$ \\
\hline
\hline
SENS approx & $30000$     & $10000$      & $4 \times 10^{11}$ \\
SENS approx & $60000$     & $10000$      & $8 \times 10^{11}$ \\
SENS exact  & $60000$     & $10000$      & $8 \times 10^{11}$  \\
\lasthline
\end{tabular}
\caption{Comparison of methods used to obtain the LJ$_{75}$ heat capacity curves shown in Fig.~\ref{fig:cv_lj75}. $N_{\text{E}}^{(total)}$ indicates the total number of energy evaluations (summed over all processors). PT curves are not shown as the computational cost to converge its heat capacity by this method is computationally prohibitive as shown in Ref.~\cite{MANDELSHTAM}. SENS exact does not converge as quickly as approximate SENS due to the low accuracy of the HSA and hence the low swap acceptance.}
\label{table:lj75}
\end{table}

\subsection*{Methods}
We define a move in MCMC as the displacement of each individual particle along
a random vector ($n$ in total). After each MCMC walk we update the step size in
order to keep the average acceptance ratio within range of some target value,
which we have chosen as $0.5$. The default parameter
values for the onset function Eq.~(\ref{eq:onset_function}) were $f_{max} \approx
2/3$ and $\Delta E=1$. We used a spherical box of radius $R=2.5~\sigma$ for
$n = 31$, $R=2.8~\sigma$ for $n = 38$ and $R=3.0~\sigma$ for $n=75$, with no periodic boundary conditions and no cutoff radius. All calculations were carried out
on a single workstation with $\mathcal{P}=16$ processors (eight-core dual Xeon
E5-2670 2.6GHz, Westmere) using the improved parallelization scheme discussed in
sec.~Parallelization. The calculations were terminated when the
energy difference between the replicas with highest and lowest energies was less
than $10^{-2}~\epsilon$. Energies of the final ``live'' replicas
were added to the output and the compression factor associated with the $\ell^{th}$ ``live'' replica was computed as
\begin{equation}
 \mu_\ell^{(live)} = \prod_{j=0}^{\ell<K}\frac{K-j}{K-j+1}.
\end{equation}
Error bars were obtained by the compression factor resampling scheme discussed in the Supplementary Information. 
By nested sampling or SENS iterations, $N_{\text{iter}}$, we mean a whole nested sampling
iteration on $\mathcal{P}$ processors, the total number of energy evaluations
is $N_{E}^{(tot)} = N \times \mathcal{P} \times N_{\text{iter}}$, where $N$ is the
number of steps in a MCMC. The computational overhead associated with global optimization by basin-hopping is insignificant as less than around $O(10^5)$ energy evaluations are necessary to find the global minima of the LJ clusters considered here~\cite{C3CP44332A}. Highly modular Python/C parallel implementations of nested sampling and SENS are publicly available~\cite{Martiniani2013, Martiniani2013a}.\\

\begin{acknowledgements}
We gratefully acknowledge discussions with Andy Ballard, Victor Ruehle, Thomas Stecher, Tine Curk, Boris Fackovec and Robert Baldock. This work has been supported by the European Research Council and the EPSRC. S.M. acknowledges financial support from the Gates Cambridge Scholarship and J.D.S. from the Marie Curie IEF 275544. D.F. acknowledges support from the ERC Advanced Grant 227758, Wolfson Merit Award 2007/R3 of the Royal Society of London. 
\end{acknowledgements}

\bibliography{newbib.bib}

\begin{thebibliography}{10}%
\makeatletter
\providecommand \@ifxundefined [1]{%
 \ifx #1\undefined \expandafter \@firstoftwo
 \else \expandafter \@secondoftwo
\fi
}%
\providecommand \@ifnum [1]{%
 \ifnum #1\expandafter \@firstoftwo
 \else \expandafter \@secondoftwo
\fi
}%
\providecommand \enquote [1]{``#1''}%
\providecommand \bibnamefont  [1]{#1}%
\providecommand \bibfnamefont [1]{#1}%
\providecommand \citenamefont [1]{#1}%
\providecommand\href[0]{\@sanitize\@href}%
\providecommand\@href[1]{\endgroup\@@startlink{#1}\endgroup\@@href}%
\providecommand\@@href[1]{#1\@@endlink}%
\providecommand \@sanitize [0]{\begingroup\catcode`\&12\catcode`\#12\relax}%
\@ifxundefined \pdfoutput {\@firstoftwo}{%
 \@ifnum{\z@=\pdfoutput}{\@firstoftwo}{\@secondoftwo}%
}{%
 \providecommand\@@startlink[1]{\leavevmode\special{html:<a href="#1">}}%
 \providecommand\@@endlink[0]{\special{html:</a>}}%
}{%
 \providecommand\@@startlink[1]{%
  \leavevmode
  \pdfstartlink
   attr{/Border[0 0 1 ]/H/I/C[0 1 1]}%
   user{/Subtype/Link/A<</Type/Action/S/URI/URI(#1)>>}%
  \relax
 }%
 \providecommand\@@endlink[0]{\pdfendlink}%
}%
\providecommand \url  [0]{\begingroup\@sanitize \@url }%
\providecommand \@url [1]{\endgroup\@href {#1}{\urlprefix}}%
\providecommand \urlprefix [0]{URL }%
\providecommand \Eprint[0]{\href }%
\@ifxundefined \urlstyle {%
  \providecommand \doi [1]{doi:\discretionary{}{}{}#1}%
}{%
  \providecommand \doi [0]{doi:\discretionary{}{}{}\begingroup
  \urlstyle{rm}\Url }%
}%
\providecommand \doibase [0]{http://dx.doi.org/}%
\providecommand \Doi[1]{\href{\doibase#1}}%
\providecommand \bibAnnote [3]{%
  \BibitemShut{#1}%
  \begin{quotation}\noindent
    \textsc{Key:}\ #2\\\textsc{Annotation:}\ #3%
  \end{quotation}%
}%
\providecommand \bibAnnoteFile [2]{%
  \IfFileExists{#2}{\bibAnnote {#1} {#2} {\input{#2}}}{}%
}%
\providecommand \typeout [0]{\immediate \write \m@ne }%
\providecommand \selectlanguage [0]{\@gobble}%
\providecommand \bibinfo [0]{\@secondoftwo}%
\providecommand \bibfield [0]{\@secondoftwo}%
\providecommand \translation [1]{[#1]}%
\providecommand \BibitemOpen[0]{}%
\providecommand \bibitemStop [0]{}%
\providecommand \bibitemNoStop [0]{.\EOS\space}%
\providecommand \EOS [0]{\spacefactor3000\relax}%
\providecommand \BibitemShut [1]{\csname bibitem#1\endcsname}%
\bibitem{Landau2005}%
  \BibitemOpen
  \bibfield{author}{%
  \bibinfo {author} {\bibfnamefont{D.}~\bibnamefont{Landau}}\ and\ \bibinfo
  {author} {\bibfnamefont{K.}~\bibnamefont{Binder}},\ }%
  \emph{\bibinfo {title} {{A Guide to Monte Carlo Simulations in Statistical
  Physics}}}\ (\bibinfo {publisher} {Cambridge University Press},\ \bibinfo
  {year} {2005})%
  \bibAnnoteFile{NoStop}{Landau2005}%
\bibitem{Wales2003}%
  \BibitemOpen
  \bibfield{author}{%
  \bibinfo {author} {\bibfnamefont{D.~J.}\ \bibnamefont{Wales}},\ }%
  \emph{\bibinfo {title} {{Energy Landscapes: Applications to Clusters,
  Biomolecules and Glasses}}}\ (\bibinfo {publisher} {Cambridge University
  Press},\ \bibinfo {year} {2005})%
  \bibAnnoteFile{NoStop}{Wales2003}%
\bibitem{Torrie1977}%
  \BibitemOpen
  \bibfield{author}{%
  \bibinfo {author} {\bibfnamefont{G.~M.}\ \bibnamefont{Torrie}}\ and\ \bibinfo
  {author} {\bibfnamefont{J.~P.}\ \bibnamefont{Valleau}},\ }%
  \bibfield{title}{%
  \enquote{\bibinfo {title} {{Nonphysical sampling distributions in Monte Carlo
  free-energy estimation: Umbrella sampling}},}\ }%
  \bibfield{journal}{%
  \bibinfo {journal} {Journal of Computational Physics}\ }%
  \textbf{\bibinfo {volume} {23}},\ \bibinfo {pages} {187 -- 199} (\bibinfo
  {year} {1977})%
  \bibAnnoteFile{NoStop}{Torrie1977}%
\bibitem{Berg1992}%
  \BibitemOpen
  \bibfield{author}{%
  \bibinfo {author} {\bibfnamefont{B.~A.}\ \bibnamefont{Berg}}\ and\ \bibinfo
  {author} {\bibfnamefont{T.}~\bibnamefont{Neuhaus}},\ }%
  \bibfield{title}{%
  \enquote{\bibinfo {title} {{Multicanonical ensemble: A new approach to
  simulate first-order phase transitions}},}\ }%
  \bibfield{journal}{%
  \bibinfo {journal} {Physical Review Letters}\ }%
  \textbf{\bibinfo {volume} {68}},\ \bibinfo {pages} {9--12} (\bibinfo {year}
  {1992})%
  \bibAnnoteFile{NoStop}{Berg1992}%
\bibitem{Wang2001}%
  \BibitemOpen
  \bibfield{author}{%
  \bibinfo {author} {\bibfnamefont{F.}~\bibnamefont{Wang}}\ and\ \bibinfo
  {author} {\bibfnamefont{D.}~\bibnamefont{Landau}},\ }%
  \bibfield{title}{%
  \enquote{\bibinfo {title} {{Efficient, Multiple-Range Random Walk Algorithm
  to Calculate the Density of States}},}\ }%
  \bibfield{journal}{%
  \bibinfo {journal} {Physical Review Letters}\ }%
  \textbf{\bibinfo {volume} {86}},\ \bibinfo {pages} {2050--2053} (\bibinfo
  {year} {2001})%
  \bibAnnoteFile{NoStop}{Wang2001}%
\bibitem{Swendsen1986}%
  \BibitemOpen
  \bibfield{author}{%
  \bibinfo {author} {\bibfnamefont{R.~H.}\ \bibnamefont{Swendsen}}\ and\
  \bibinfo {author} {\bibfnamefont{J.-S.}\ \bibnamefont{Wang}},\ }%
  \bibfield{title}{%
  \enquote{\bibinfo {title} {{Replica Monte Carlo Simulation of
  Spin-Glasses}},}\ }%
  \bibfield{journal}{%
  \bibinfo {journal} {Physical Review Letters}\ }%
  \textbf{\bibinfo {volume} {57}},\ \bibinfo {pages} {2607--2609} (\bibinfo
  {year} {1986})%
  \bibAnnoteFile{NoStop}{Swendsen1986}%
\bibitem{Swendsen1987}%
  \BibitemOpen
  \bibfield{author}{%
  \bibinfo {author} {\bibfnamefont{F.}~\bibnamefont{Wang}}\ and\ \bibinfo
  {author} {\bibfnamefont{D.}~\bibnamefont{Landau}},\ }%
  \bibfield{title}{%
  \enquote{\bibinfo {title} {{Efficient, Multiple-Range Random Walk Algorithm
  to Calculate the Density of States}},}\ }%
  \bibfield{journal}{%
  \bibinfo {journal} {Physical Review Letters}\ }%
  \textbf{\bibinfo {volume} {86}},\ \bibinfo {pages} {2050--2053} (\bibinfo
  {year} {2001})%
  \bibAnnoteFile{NoStop}{Swendsen1987}%
\bibitem{Sugita1999}%
  \BibitemOpen
  \bibfield{author}{%
  \bibinfo {author} {\bibfnamefont{Y.}~\bibnamefont{Sugita}}\ and\ \bibinfo
  {author} {\bibfnamefont{Y.}~\bibnamefont{Okamoto}},\ }%
  \bibfield{title}{%
  \enquote{\bibinfo {title} {{Replica-exchange molecular dynamics method for
  protein folding}},}\ }%
  \bibfield{journal}{%
  \bibinfo {journal} {Chemical Physics Letters}\ }%
  \textbf{\bibinfo {volume} {314}},\ \bibinfo {pages} {141--151} (\bibinfo
  {year} {1999})%
  \bibAnnoteFile{NoStop}{Sugita1999}%
\bibitem{Laio2002}%
  \BibitemOpen
  \bibfield{author}{%
  \bibinfo {author} {\bibfnamefont{A.}~\bibnamefont{Laio}}\ and\ \bibinfo
  {author} {\bibfnamefont{M.}~\bibnamefont{Parrinello}},\ }%
  \bibfield{title}{%
  \enquote{\bibinfo {title} {{Escaping free-energy minima}},}\ }%
  \bibfield{journal}{%
  \bibinfo {journal} {Proceedings of the National Academy of Sciences}\ }%
  \textbf{\bibinfo {volume} {99}},\ \bibinfo {pages} {12562--12566} (\bibinfo
  {year} {2002})%
  \bibAnnoteFile{NoStop}{Laio2002}%
\bibitem{Ferrenberg1988}%
  \BibitemOpen
  \bibfield{author}{%
  \bibinfo {author} {\bibfnamefont{A.}~\bibnamefont{Ferrenberg}}\ and\ \bibinfo
  {author} {\bibfnamefont{R.}~\bibnamefont{Swendsen}},\ }%
  \bibfield{title}{%
  \enquote{\bibinfo {title} {{New Monte Carlo technique for studying phase
  transitions}},}\ }%
  \bibfield{journal}{%
  \bibinfo {journal} {Physical Review Letters}\ }%
  \textbf{\bibinfo {volume} {61}},\ \bibinfo {pages} {2635--2638} (\bibinfo
  {year} {1988})%
  \bibAnnoteFile{NoStop}{Ferrenberg1988}%
\bibitem{Chodera2007}%
  \BibitemOpen
  \bibfield{author}{%
  \bibinfo {author} {\bibfnamefont{J.~D.}\ \bibnamefont{Chodera}}, \bibinfo
  {author} {\bibfnamefont{W.~C.}\ \bibnamefont{Swope}}, \bibinfo {author}
  {\bibfnamefont{J.~W.}\ \bibnamefont{Pitera}}, \bibinfo {author}
  {\bibfnamefont{C.}~\bibnamefont{Seok}},\ and\ \bibinfo {author}
  {\bibfnamefont{K.~A.}\ \bibnamefont{Dill}},\ }%
  \bibfield{title}{%
  \enquote{\bibinfo {title} {{Use of the Weighted Histogram Analysis Method for
  the Analysis of Simulated and Parallel Tempering Simulations}},}\ }%
  \bibfield{journal}{%
  \bibinfo {journal} {Journal of Chemical Theory and Computation}\ }%
  \textbf{\bibinfo {volume} {3}},\ \bibinfo {pages} {26--41} (\bibinfo {year}
  {2007})%
  \bibAnnoteFile{NoStop}{Chodera2007}%
\bibitem{Wales2013}%
  \BibitemOpen
  \bibfield{author}{%
  \bibinfo {author} {\bibfnamefont{D.~J.}\ \bibnamefont{Wales}},\ }%
  \bibfield{title}{%
  \enquote{\bibinfo {title} {{Surveying a complex potential energy landscape:
  Overcoming broken ergodicity using basin-sampling}},}\ }%
  \bibfield{journal}{%
  \bibinfo {journal} {Chemical Physics Letters}\ }%
  \textbf{\bibinfo {volume} {584}},\ \bibinfo {pages} {1--9} (\bibinfo {year}
  {2013})%
  \bibAnnoteFile{NoStop}{Wales2013}%
\bibitem{Doye1999}%
  \BibitemOpen
  \bibfield{author}{%
  \bibinfo {author} {\bibfnamefont{J.~P.~K.}\ \bibnamefont{Doye}}, \bibinfo
  {author} {\bibfnamefont{M.~A.}\ \bibnamefont{Miller}},\ and\ \bibinfo
  {author} {\bibfnamefont{D.~J.}\ \bibnamefont{Wales}},\ }%
  \bibfield{title}{%
  \enquote{\bibinfo {title} {{Evolution of the potential energy surface with
  size for Lennard-Jones clusters}},}\ }%
  \bibfield{journal}{%
  \bibinfo {journal} {The Journal of Chemical Physics}\ }%
  \textbf{\bibinfo {volume} {111}},\ \bibinfo {pages} {8417} (\bibinfo {year}
  {1999})%
  \bibAnnoteFile{NoStop}{Doye1999}%
\bibitem{Doye1998}%
  \BibitemOpen
  \bibfield{author}{%
  \bibinfo {author} {\bibfnamefont{J.~P.~K.}\ \bibnamefont{Doye}}\ and\
  \bibinfo {author} {\bibfnamefont{D.~J.}\ \bibnamefont{Wales}},\ }%
  \bibfield{title}{%
  \enquote{\bibinfo {title} {{Thermodynamics of Global Optimization}},}\ }%
  \bibfield{journal}{%
  \bibinfo {journal} {Physical Review Letters}\ }%
  \textbf{\bibinfo {volume} {80}},\ \bibinfo {pages} {1357--1360} (\bibinfo
  {year} {1998})%
  \bibAnnoteFile{NoStop}{Doye1998}%
\bibitem{Poulain2006}%
  \BibitemOpen
  \bibfield{author}{%
  \bibinfo {author} {\bibfnamefont{P.}~\bibnamefont{Poulain}}, \bibinfo
  {author} {\bibfnamefont{F.}~\bibnamefont{Calvo}}, \bibinfo {author}
  {\bibfnamefont{R.}~\bibnamefont{Antoine}}, \bibinfo {author}
  {\bibfnamefont{M.}~\bibnamefont{Broyer}},\ and\ \bibinfo {author}
  {\bibfnamefont{Ph.}\ \bibnamefont{Dugourd}},\ }%
  \bibfield{title}{%
  \enquote{\bibinfo {title} {{Performances of Wang-Landau algorithms for
  continuous systems}},}\ }%
  \bibfield{journal}{%
  \bibinfo {journal} {Physical Review E}\ }%
  \textbf{\bibinfo {volume} {73}},\ \bibinfo {pages} {056704} (\bibinfo {year}
  {2006})%
  \bibAnnoteFile{NoStop}{Poulain2006}%
\bibitem{MANDELSHTAM}%
  \BibitemOpen
  \bibfield{author}{%
  \bibinfo {author} {\bibfnamefont{V.~A.}\ \bibnamefont{Mandelshtam}}, \bibinfo
  {author} {\bibfnamefont{P.~A.}\ \bibnamefont{Frantsuzov}},\ and\ \bibinfo
  {author} {\bibfnamefont{F.}~\bibnamefont{Calvo}},\ }%
  \bibfield{title}{%
  \enquote{\bibinfo {title} {{Structural transitions and melting in LJ74-78
  lennard-jones clusters from adaptive exchange monte carlo simulations}},}\ }%
  \bibfield{journal}{%
  \bibinfo {journal} {The Journal of Physical Chemistry A}\ }%
  \textbf{\bibinfo {volume} {110}},\ \bibinfo {pages} {5326--5332} (\bibinfo
  {year} {2007})%
  \bibAnnoteFile{NoStop}{MANDELSHTAM}%
\bibitem{SHARAPOV}%
  \BibitemOpen
  \bibfield{author}{%
  \bibinfo {author} {\bibfnamefont{V.~A.}\ \bibnamefont{Sharapov}}\ and\
  \bibinfo {author} {\bibfnamefont{V.~A.}\ \bibnamefont{Mandelshtam}},\ }%
  \bibfield{title}{%
  \enquote{\bibinfo {title} {{Solid-solid structural transformations in
  lennard-jones clusters : Accurate simulations versus the harmonic
  superposition approximation}},}\ }%
  \bibfield{journal}{%
  \bibinfo {journal} {The Journal of Physical Chemistry A}\ }%
  \textbf{\bibinfo {volume} {111}},\ \bibinfo {pages} {10284--10291} (\bibinfo
  {year} {2007})%
  \bibAnnoteFile{NoStop}{SHARAPOV}%
\bibitem{Sharapov2007}%
  \BibitemOpen
  \bibfield{author}{%
  \bibinfo {author} {\bibfnamefont{V.~A.}\ \bibnamefont{Sharapov}}, \bibinfo
  {author} {\bibfnamefont{D.}~\bibnamefont{Meluzzi}},\ and\ \bibinfo {author}
  {\bibfnamefont{V.~A.}\ \bibnamefont{Mandelshtam}},\ }%
  \bibfield{title}{%
  \enquote{\bibinfo {title} {{Low-Temperature Structural Transitions:
  Circumventing the Broken-Ergodicity Problem}},}\ }%
  \bibfield{journal}{%
  \bibinfo {journal} {Physical Review Letters}\ }%
  \textbf{\bibinfo {volume} {98}},\ \bibinfo {pages} {105701} (\bibinfo {year}
  {2007})%
  \bibAnnoteFile{NoStop}{Sharapov2007}%
\bibitem{Skilling2006}%
  \BibitemOpen
  \bibfield{author}{%
  \bibinfo {author} {\bibfnamefont{J.}~\bibnamefont{Skilling}},\ }%
  \bibfield{title}{%
  \enquote{\bibinfo {title} {{Nested sampling for general Bayesian
  computation}},}\ }%
  \bibfield{journal}{%
  \bibinfo {journal} {Bayesian Analysis}\ }%
  \textbf{\bibinfo {volume} {1}},\ \bibinfo {pages} {833--859} (\bibinfo {year}
  {2006})%
  \bibAnnoteFile{NoStop}{Skilling2006}%
\bibitem{Mukherjee2006a}%
  \BibitemOpen
  \bibfield{author}{%
  \bibinfo {author} {\bibfnamefont{P.}~\bibnamefont{Mukherjee}}, \bibinfo
  {author} {\bibfnamefont{D.}~\bibnamefont{Parkinson}},\ and\ \bibinfo {author}
  {\bibfnamefont{A.~R.}\ \bibnamefont{Liddle}},\ }%
  \bibfield{title}{%
  \enquote{\bibinfo {title} {{A Nested Sampling Algorithm for Cosmological
  Model Selection}},}\ }%
  \bibfield{journal}{%
  \bibinfo {journal} {The Astrophysical Journal}\ }%
  \textbf{\bibinfo {volume} {638}},\ \bibinfo {pages} {L51--L54} (\bibinfo
  {year} {2006})%
  \bibAnnoteFile{NoStop}{Mukherjee2006a}%
\bibitem{Feroz2008}%
  \BibitemOpen
  \bibfield{author}{%
  \bibinfo {author} {\bibfnamefont{F.}~\bibnamefont{Feroz}}\ and\ \bibinfo
  {author} {\bibfnamefont{M.~P.}\ \bibnamefont{Hobson}},\ }%
  \bibfield{title}{%
  \enquote{\bibinfo {title} {{Multimodal nested sampling: an efficient and
  robust alternative to Markov Chain Monte Carlo methods for astronomical data
  analyses}},}\ }%
  \bibfield{journal}{%
  \bibinfo {journal} {Monthly Notices of the Royal Astronomical Society}\ }%
  \textbf{\bibinfo {volume} {384}},\ \bibinfo {pages} {449--463} (\bibinfo
  {year} {2008})%
  \bibAnnoteFile{NoStop}{Feroz2008}%
\bibitem{Shaw2007a}%
  \BibitemOpen
  \bibfield{author}{%
  \bibinfo {author} {\bibfnamefont{J.~R.}\ \bibnamefont{Shaw}}, \bibinfo
  {author} {\bibfnamefont{M.}~\bibnamefont{Bridges}},\ and\ \bibinfo {author}
  {\bibfnamefont{M.~P.}\ \bibnamefont{Hobson}},\ }%
  \bibfield{title}{%
  \enquote{\bibinfo {title} {{Efficient Bayesian inference for multimodal
  problems in cosmology}},}\ }%
  \bibfield{journal}{%
  \bibinfo {journal} {Monthly Notices of the Royal Astronomical Society}\ }%
  \textbf{\bibinfo {volume} {378}},\ \bibinfo {pages} {1365--1370} (\bibinfo
  {year} {2007})%
  \bibAnnoteFile{NoStop}{Shaw2007a}%
\bibitem{Feroz2009}%
  \BibitemOpen
  \bibfield{author}{%
  \bibinfo {author} {\bibfnamefont{F.}~\bibnamefont{Feroz}}, \bibinfo {author}
  {\bibfnamefont{M.~P.}\ \bibnamefont{Hobson}},\ and\ \bibinfo {author}
  {\bibfnamefont{M.}~\bibnamefont{Bridges}},\ }%
  \bibfield{title}{%
  \enquote{\bibinfo {title} {{MultiNest: an efficient and robust Bayesian
  inference tool for cosmology and particle physics}},}\ }%
  \bibfield{journal}{%
  \bibinfo {journal} {Monthly Notices of the Royal Astronomical Society}\ }%
  \textbf{\bibinfo {volume} {398}},\ \bibinfo {pages} {1601--1614} (\bibinfo
  {year} {2009})%
  \bibAnnoteFile{NoStop}{Feroz2009}%
\bibitem{murray2006}%
  \BibitemOpen
  \bibfield{author}{%
  \bibinfo {author} {\bibfnamefont{I.}~\bibnamefont{Murray}}, \bibinfo {author}
  {\bibfnamefont{D.~J.~C.}\ \bibnamefont{MacKay}}, \bibinfo {author}
  {\bibfnamefont{Z.}~\bibnamefont{Ghahramani}},\ and\ \bibinfo {author}
  {\bibfnamefont{J.}~\bibnamefont{Skilling}},\ }%
  \bibfield{title}{%
  \enquote{\bibinfo {title} {{Nested sampling for Potts models}},}\ }%
  \bibfield{journal}{%
  \bibinfo {journal} {Advances in Neural Information Processing Systems}\ }%
  \textbf{\bibinfo {volume} {18}},\ \bibinfo {pages} {947} (\bibinfo {year}
  {2006})%
  \bibAnnoteFile{NoStop}{murray2006}%
\bibitem{Partay2010}%
  \BibitemOpen
  \bibfield{author}{%
  \bibinfo {author} {\bibfnamefont{L.~B.}\ \bibnamefont{P\'{a}rtay}}, \bibinfo
  {author} {\bibfnamefont{A.~P.}\ \bibnamefont{Bart\'{o}k}},\ and\ \bibinfo
  {author} {\bibfnamefont{G.}~\bibnamefont{Cs\'{a}nyi}},\ }%
  \bibfield{title}{%
  \enquote{\bibinfo {title} {{Efficient sampling of atomic configurational
  spaces}},}\ }%
  \bibfield{journal}{%
  \bibinfo {journal} {The Journal of Physical Chemistry B}\ }%
  \textbf{\bibinfo {volume} {114}},\ \bibinfo {pages} {10502--12} (\bibinfo
  {year} {2010})%
  \bibAnnoteFile{NoStop}{Partay2010}%
\bibitem{Nielsen2013}%
  \BibitemOpen
  \bibfield{author}{%
  \bibinfo {author} {\bibfnamefont{S.~O.}\ \bibnamefont{Nielsen}},\ }%
  \bibfield{title}{%
  \enquote{\bibinfo {title} {{Nested sampling in the canonical ensemble: Direct
  calculation of the partition function from NVT trajectories}},}\ }%
  \bibfield{journal}{%
  \bibinfo {journal} {The Journal of Chemical Physics}\ }%
  \textbf{\bibinfo {volume} {139}},\ \bibinfo {pages} {124104} (\bibinfo {year}
  {2013})%
  \bibAnnoteFile{NoStop}{Nielsen2013}%
\bibitem{Burkoff2012}%
  \BibitemOpen
  \bibfield{author}{%
  \bibinfo {author} {\bibfnamefont{N.~S.}\ \bibnamefont{Burkoff}}, \bibinfo
  {author} {\bibfnamefont{C.}~\bibnamefont{V\'{a}rnai}}, \bibinfo {author}
  {\bibfnamefont{S.~A.}\ \bibnamefont{Wells}},\ and\ \bibinfo {author}
  {\bibfnamefont{D.~L.}\ \bibnamefont{Wild}},\ }%
  \bibfield{title}{%
  \enquote{\bibinfo {title} {{Exploring the energy landscapes of protein
  folding simulations with Bayesian computation}},}\ }%
  \bibfield{journal}{%
  \bibinfo {journal} {Biophysical Journal}\ }%
  \textbf{\bibinfo {volume} {102}},\ \bibinfo {pages} {878--86} (\bibinfo
  {year} {2012})%
  \bibAnnoteFile{NoStop}{Burkoff2012}%
\bibitem{Partay2012}%
  \BibitemOpen
  \bibfield{author}{%
  \bibinfo {author} {\bibfnamefont{L.~B.}\ \bibnamefont{P\'{a}rtay}}, \bibinfo
  {author} {\bibfnamefont{A.~P.}\ \bibnamefont{Bart\'{o}k}},\ and\ \bibinfo
  {author} {\bibfnamefont{G.}~\bibnamefont{Cs\'{a}nyi}},\ }%
  \bibfield{title}{%
  \enquote{\bibinfo {title} {{Nested sampling for materials: The case of hard
  spheres}},}\ }%
  \bibfield{journal}{%
  \bibinfo {journal} {Physical Review E}\ }%
  \textbf{\bibinfo {volume} {89}},\ \bibinfo {pages} {022302} (\bibinfo {year}
  {2014})%
  \bibAnnoteFile{NoStop}{Partay2012}%
\bibitem{Brewer2010}%
  \BibitemOpen
  \bibfield{author}{%
  \bibinfo {author} {\bibfnamefont{B.~J.}\ \bibnamefont{Brewer}}, \bibinfo
  {author} {\bibfnamefont{L.~B.}\ \bibnamefont{P\'{a}rtay}},\ and\ \bibinfo
  {author} {\bibfnamefont{G.}~\bibnamefont{Cs\'{a}nyi}},\ }%
  \bibfield{title}{%
  \enquote{\bibinfo {title} {{Diffusive nested sampling}},}\ }%
  \bibfield{journal}{%
  \bibinfo {journal} {Statistics and Computing}\ }%
  \textbf{\bibinfo {volume} {21}},\ \bibinfo {pages} {649--656} (\bibinfo
  {year} {2010})%
  \bibAnnoteFile{NoStop}{Brewer2010}%
\bibitem{Aitken2013}%
  \BibitemOpen
  \bibfield{author}{%
  \bibinfo {author} {\bibfnamefont{S.}~\bibnamefont{Aitken}}\ and\ \bibinfo
  {author} {\bibfnamefont{O.}~\bibnamefont{Akman}},\ }%
  \bibfield{title}{%
  \enquote{\bibinfo {title} {{Nested sampling for parameter inference in
  systems biology: application to an exemplar circadian model}},}\ }%
  \bibfield{journal}{%
  \bibinfo {journal} {BMC Systems Biology}\ }%
  \textbf{\bibinfo {volume} {7}} (\bibinfo {year} {2013})%
  \bibAnnoteFile{NoStop}{Aitken2013}%
\bibitem{Dybowski2013}%
  \BibitemOpen
  \bibfield{author}{%
  \bibinfo {author} {\bibfnamefont{R.}~\bibnamefont{Dybowski}}, \bibinfo
  {author} {\bibfnamefont{T.~J.}\ \bibnamefont{McKinley}}, \bibinfo {author}
  {\bibfnamefont{P.}~\bibnamefont{Mastroeni}},\ and\ \bibinfo {author}
  {\bibfnamefont{O.}~\bibnamefont{Restif}},\ }%
  \bibfield{title}{%
  \enquote{\bibinfo {title} {{Nested sampling for bayesian model comparison in
  the context of salmonella disease dynamics}},}\ }%
  \bibfield{journal}{%
  \bibinfo {journal} {PloS one}\ }%
  \textbf{\bibinfo {volume} {8}},\ \bibinfo {pages} {e82317} (\bibinfo {year}
  {2013})%
  \bibAnnoteFile{NoStop}{Dybowski2013}%
\bibitem{Pullen2014}%
  \BibitemOpen
  \bibfield{author}{%
  \bibinfo {author} {\bibfnamefont{N.}~\bibnamefont{Pullen}}\ and\ \bibinfo
  {author} {\bibfnamefont{R.~J.}\ \bibnamefont{Morris}},\ }%
  \bibfield{title}{%
  \enquote{\bibinfo {title} {{Bayesian model comparison and parameter inference
  in systems biology using nested sampling}},}\ }%
  \bibfield{journal}{%
  \bibinfo {journal} {PloS one}\ }%
  \textbf{\bibinfo {volume} {9}},\ \bibinfo {pages} {e88419} (\bibinfo {year}
  {2014})%
  \bibAnnoteFile{NoStop}{Pullen2014}%
\bibitem{Li1987}%
  \BibitemOpen
  \bibfield{author}{%
  \bibinfo {author} {\bibfnamefont{Z.}~\bibnamefont{Li}}\ and\ \bibinfo
  {author} {\bibfnamefont{H.~A.}\ \bibnamefont{Scheraga}},\ }%
  \bibfield{title}{%
  \enquote{\bibinfo {title} {{Monte Carlo-minimization approach to the
  multiple-minima problem in protein folding}},}\ }%
  \bibfield{journal}{%
  \bibinfo {journal} {Proceedings of the National Academy of Sciences of the
  United States of America}\ }%
  \textbf{\bibinfo {volume} {84}},\ \bibinfo {pages} {6611--5} (\bibinfo {year}
  {1987})%
  \bibAnnoteFile{NoStop}{Li1987}%
\bibitem{Wales1997}%
  \BibitemOpen
  \bibfield{author}{%
  \bibinfo {author} {\bibfnamefont{D.~J.}\ \bibnamefont{Wales}}\ and\ \bibinfo
  {author} {\bibfnamefont{J.~P.~K.}\ \bibnamefont{Doye}},\ }%
  \bibfield{title}{%
  \enquote{\bibinfo {title} {{Global Optimization by Basin-Hopping and the
  Lowest Energy Structures of Lennard-Jones Clusters Containing up to 110
  Atoms}},}\ }%
  \bibfield{journal}{%
  \bibinfo {journal} {The Journal of Physical Chemistry A}\ }%
  \textbf{\bibinfo {volume} {101}},\ \bibinfo {pages} {5111--5116} (\bibinfo
  {year} {1997})%
  \bibAnnoteFile{NoStop}{Wales1997}%
\bibitem{C3CP44332A}%
  \BibitemOpen
  \bibfield{author}{%
  \bibinfo {author} {\bibfnamefont{M.~T.}\ \bibnamefont{Oakley}}, \bibinfo
  {author} {\bibfnamefont{R.~L.}\ \bibnamefont{Johnston}},\ and\ \bibinfo
  {author} {\bibfnamefont{D.~J.}\ \bibnamefont{Wales}},\ }%
  \bibfield{title}{%
  \enquote{\bibinfo {title} {{Symmetrisation schemes for global optimisation of
  atomic clusters}},}\ }%
  \bibfield{journal}{%
  \bibinfo {journal} {Physical Chemistry Chemical Physics}\ }%
  \textbf{\bibinfo {volume} {15}},\ \bibinfo {pages} {3965--3976} (\bibinfo
  {year} {2013})%
  \bibAnnoteFile{NoStop}{C3CP44332A}%
\bibitem{mcginty71}%
  \BibitemOpen
  \bibfield{author}{%
  \bibinfo {author} {\bibfnamefont{D.~J.}\ \bibnamefont{McGinty}},\ }%
  \bibfield{title}{%
  \enquote{\bibinfo {title} {{Vapor phase homogenous nucleation and the
  thermodynamic properties of small clusters of argon atoms}},}\ }%
  \bibfield{journal}{%
  \bibinfo {journal} {The Journal of Chemical Physics}\ }%
  \textbf{\bibinfo {volume} {55}},\ \bibinfo {pages} {580} (\bibinfo {year}
  {1971})%
  \bibAnnoteFile{NoStop}{mcginty71}%
\bibitem{burton72}%
  \BibitemOpen
  \bibfield{author}{%
  \bibinfo {author} {\bibfnamefont{J.~J.}\ \bibnamefont{Burton}},\ }%
  \bibfield{title}{%
  \enquote{\bibinfo {title} {{Vibrational Frequencies and Entropies of Small
  Clusters of Atoms}},}\ }%
  \bibfield{journal}{%
  \bibinfo {journal} {The Journal of Chemical Physics}\ }%
  \textbf{\bibinfo {volume} {56}},\ \bibinfo {pages} {3133} (\bibinfo {year}
  {1972})%
  \bibAnnoteFile{NoStop}{burton72}%
\bibitem{hoare79}%
  \BibitemOpen
  \bibfield{author}{%
  \bibinfo {author} {\bibfnamefont{M.~R.}\ \bibnamefont{Hoare}},\ }%
  \bibfield{title}{%
  \enquote{\bibinfo {title} {{Structure and dynamics of simple
  microclusters}},}\ }%
  \bibfield{journal}{%
  \bibinfo {journal} {Advances in Chemical Physics}\ }%
  \textbf{\bibinfo {volume} {40}},\ \bibinfo {pages} {49} (\bibinfo {year}
  {1979})%
  \bibAnnoteFile{NoStop}{hoare79}%
\bibitem{stillingerw84}%
  \BibitemOpen
  \bibfield{author}{%
  \bibinfo {author} {\bibfnamefont{F.~H.}\ \bibnamefont{Stillinger}}\ and\
  \bibinfo {author} {\bibfnamefont{T.~A.}\ \bibnamefont{Weber}},\ }%
  \bibfield{title}{%
  \enquote{\bibinfo {title} {packing structures and transitions in liquids and
  solids},}\ }%
  \bibfield{journal}{%
  \bibinfo {journal} {Science}\ }%
  \textbf{\bibinfo {volume} {225}},\ \bibinfo {pages} {983} (\bibinfo {year}
  {1984})%
  \bibAnnoteFile{NoStop}{stillingerw84}%
\bibitem{Strodel2008}%
  \BibitemOpen
  \bibfield{author}{%
  \bibinfo {author} {\bibfnamefont{B.}~\bibnamefont{Strodel}}\ and\ \bibinfo
  {author} {\bibfnamefont{D.~J.}\ \bibnamefont{Wales}},\ }%
  \bibfield{title}{%
  \enquote{\bibinfo {title} {{Free energy surfaces from an extended harmonic
  superposition approach and kinetics for alanine dipeptide}},}\ }%
  \bibfield{journal}{%
  \bibinfo {journal} {Chemical Physics Letters}\ }%
  \textbf{\bibinfo {volume} {466}},\ \bibinfo {pages} {105--115} (\bibinfo
  {year} {2008})%
  \bibAnnoteFile{NoStop}{Strodel2008}%
\bibitem{Wales1993}%
  \BibitemOpen
  \bibfield{author}{%
  \bibinfo {author} {\bibfnamefont{D.~J.}\ \bibnamefont{Wales}},\ }%
  \bibfield{title}{%
  \enquote{\bibinfo {title} {{Coexistence in small inert gas clusters}},}\ }%
  \bibfield{journal}{%
  \bibinfo {journal} {Molecular Physics}\ }%
  \textbf{\bibinfo {volume} {78}},\ \bibinfo {pages} {151--171} (\bibinfo
  {year} {1993})%
  \bibAnnoteFile{NoStop}{Wales1993}%
\bibitem{Doye1995}%
  \BibitemOpen
  \bibfield{author}{%
  \bibinfo {author} {\bibfnamefont{J.~P.~K.}\ \bibnamefont{Doye}}\ and\
  \bibinfo {author} {\bibfnamefont{D.~J.}\ \bibnamefont{Wales}},\ }%
  \bibfield{title}{%
  \enquote{\bibinfo {title} {{An order parameter approach to coexistence in
  atomic clusters}},}\ }%
  \bibfield{journal}{%
  \bibinfo {journal} {The Journal of Chemical Physics}\ }%
  \textbf{\bibinfo {volume} {102}},\ \bibinfo {pages} {9673} (\bibinfo {year}
  {1995})%
  \bibAnnoteFile{NoStop}{Doye1995}%
\bibitem{Doye1995a}%
  \BibitemOpen
  \bibfield{author}{%
  \bibinfo {author} {\bibfnamefont{J.~P.~K.}\ \bibnamefont{Doye}}\ and\
  \bibinfo {author} {\bibfnamefont{D.~J.}\ \bibnamefont{Wales}},\ }%
  \bibfield{title}{%
  \enquote{\bibinfo {title} {{Calculation of thermodynamic properties of small
  Lennard-Jones clusters incorporating anharmonicity}},}\ }%
  \bibfield{journal}{%
  \bibinfo {journal} {The Journal of Chemical Physics}\ }%
  \textbf{\bibinfo {volume} {102}},\ \bibinfo {pages} {9659} (\bibinfo {year}
  {1995})%
  \bibAnnoteFile{NoStop}{Doye1995a}%
\bibitem{CalvoDW01}%
  \BibitemOpen
  \bibfield{author}{%
  \bibinfo {author} {\bibfnamefont{F.}~\bibnamefont{Calvo}}, \bibinfo {author}
  {\bibfnamefont{J.~P.~K.}\ \bibnamefont{Doye}},\ and\ \bibinfo {author}
  {\bibfnamefont{D.~J.}\ \bibnamefont{Wales}},\ }%
  \bibfield{title}{%
  \enquote{\bibinfo {title} {{Characterization of Anharmonicities on Complex
  Potential Energy Surfaces: Perturbation Theory and Simulation}},}\ }%
  \bibfield{journal}{%
  \bibinfo {journal} {The Journal of Chemical Physics}\ }%
  \textbf{\bibinfo {volume} {115}},\ \bibinfo {pages} {9627--9636} (\bibinfo
  {year} {2001})%
  \bibAnnoteFile{NoStop}{CalvoDW01}%
\bibitem{georgescu:144106}%
  \BibitemOpen
  \bibfield{author}{%
  \bibinfo {author} {\bibfnamefont{I.}~\bibnamefont{Georgescu}}\ and\ \bibinfo
  {author} {\bibfnamefont{V.~A.}\ \bibnamefont{Mandelshtam}},\ }%
  \bibfield{title}{%
  \enquote{\bibinfo {title} {{Self-consistent phonons revisited. I. The role of
  thermal versus quantum fluctuations on structural transitions in large
  Lennard-Jones clusters}},}\ }%
  \bibfield{journal}{%
  \bibinfo {journal} {The Journal of Chemical Physics}\ }%
  \textbf{\bibinfo {volume} {137}},\ \bibinfo {pages} {144106} (\bibinfo {year}
  {2012})%
  \bibAnnoteFile{NoStop}{georgescu:144106}%
\bibitem{Note1}%
  \BibitemOpen
  \bibinfo {note} {The normalised fraction of configuration space $X_i= \Omega
  _{E\leq E_{\protect \mathcal {R}}}/\Omega _{tot} \in (0,1)$. Since Nested
  Sampling assumes uniform sampling in phase space, estimating $X_i$ becomes
  analogous to estimating the $\protect \mathcal {R}$-th largest position out
  of $K$ points sampled at random on a unit segment. The point with $\protect
  \mathcal {R}$-th largest position will be distributed on the unit line as
  $\protect \text {Beta}(K-\protect \mathcal {R}+1,\protect \mathcal {R})$.}%
  \bibAnnoteFile{Stop}{Note1}%
\bibitem{Hartke2011}%
  \BibitemOpen
  \bibfield{author}{%
  \bibinfo {author} {\bibfnamefont{B.}~\bibnamefont{Hartke}},\ }%
  \bibfield{title}{%
  \enquote{\bibinfo {title} {{Global optimization}},}\ }%
  \bibfield{journal}{%
  \bibinfo {journal} {Wiley Interdisciplinary Reviews: Computational Molecular
  Science}\ }%
  \textbf{\bibinfo {volume} {1}},\ \bibinfo {pages} {879--887} (\bibinfo {year}
  {2011})%
  \bibAnnoteFile{NoStop}{Hartke2011}%
\bibitem{Wales1999}%
  \BibitemOpen
  \bibfield{author}{%
  \bibinfo {author} {\bibfnamefont{D.~J.}\ \bibnamefont{Wales}},\ }%
  \bibfield{title}{%
  \enquote{\bibinfo {title} {{Global Optimization of Clusters, Crystals, and
  Biomolecules}},}\ }%
  \bibfield{journal}{%
  \bibinfo {journal} {Science}\ }%
  \textbf{\bibinfo {volume} {285}},\ \bibinfo {pages} {1368--1372} (\bibinfo
  {year} {1999})%
  \bibAnnoteFile{NoStop}{Wales1999}%
\bibitem{Stillinger1984}%
  \BibitemOpen
  \bibfield{author}{%
  \bibinfo {author} {\bibfnamefont{F.~H.}\ \bibnamefont{Stillinger}}\ and\
  \bibinfo {author} {\bibfnamefont{T.~A.}\ \bibnamefont{Weber}},\ }%
  \bibfield{title}{%
  \enquote{\bibinfo {title} {{Packing structures and transitions in liquids and
  solids}},}\ }%
  \bibfield{journal}{%
  \bibinfo {journal} {Science}\ }%
  \textbf{\bibinfo {volume} {225}},\ \bibinfo {pages} {983--9} (\bibinfo {year}
  {1984})%
  \bibAnnoteFile{NoStop}{Stillinger1984}%
\bibitem{Somani2013}%
  \BibitemOpen
  \bibfield{author}{%
  \bibinfo {author} {\bibfnamefont{S.}~\bibnamefont{Somani}}\ and\ \bibinfo
  {author} {\bibfnamefont{D.~J.}\ \bibnamefont{Wales}},\ }%
  \bibfield{title}{%
  \enquote{\bibinfo {title} {{Energy landscapes and global thermodynamics for
  alanine peptides}},}\ }%
  \bibfield{journal}{%
  \bibinfo {journal} {The Journal of Chemical Physics}\ }%
  \textbf{\bibinfo {volume} {139}},\ \bibinfo {pages} {121909} (\bibinfo {year}
  {2013})%
  \bibAnnoteFile{NoStop}{Somani2013}%
\bibitem{WalesD03}%
  \BibitemOpen
  \bibfield{author}{%
  \bibinfo {author} {\bibfnamefont{D.~J.}\ \bibnamefont{Wales}}\ and\ \bibinfo
  {author} {\bibfnamefont{J.~P.~K.}\ \bibnamefont{Doye}},\ }%
  \bibfield{title}{%
  \enquote{\bibinfo {title} {{Stationary points and dynamics in
  high-dimensional systems}},}\ }%
  \bibfield{journal}{%
  \bibinfo {journal} {The Journal of Chemical Physics}\ }%
  \textbf{\bibinfo {volume} {119}},\ \bibinfo {pages} {12409--12416} (\bibinfo
  {year} {2003})%
  \bibAnnoteFile{NoStop}{WalesD03}%
\bibitem{Lennard-Jones1931}%
  \BibitemOpen
  \bibfield{author}{%
  \bibinfo {author} {\bibfnamefont{J.~E.}\ \bibnamefont{Lennard-Jones}},\ }%
  \bibfield{title}{%
  \enquote{\bibinfo {title} {{Cohesion}},}\ }%
  \bibfield{journal}{%
  \bibinfo {journal} {Proceedings of the Physical Society}\ }%
  \textbf{\bibinfo {volume} {43}},\ \bibinfo {pages} {461--482} (\bibinfo
  {year} {1931})%
  \bibAnnoteFile{NoStop}{Lennard-Jones1931}%
\bibitem{Zhou1997}%
  \BibitemOpen
  \bibfield{author}{%
  \bibinfo {author} {\bibfnamefont{R.}~\bibnamefont{Zhou}}\ and\ \bibinfo
  {author} {\bibfnamefont{B.~J.}\ \bibnamefont{Berne}},\ }%
  \bibfield{title}{%
  \enquote{\bibinfo {title} {{Smart walking: A new method for Boltzmann
  sampling of protein conformations}},}\ }%
  \bibfield{journal}{%
  \bibinfo {journal} {The Journal of Chemical Physics}\ }%
  \textbf{\bibinfo {volume} {107}},\ \bibinfo {pages} {9185--9196} (\bibinfo
  {year} {1997})%
  \bibAnnoteFile{NoStop}{Zhou1997}%
\bibitem{Ioan2001}%
  \BibitemOpen
  \bibfield{author}{%
  \bibinfo {author} {\bibfnamefont{I.}~\bibnamefont{Andricioaei}}, \bibinfo
  {author} {\bibfnamefont{J.~E.}\ \bibnamefont{Straub}},\ and\ \bibinfo
  {author} {\bibfnamefont{A.~F.}\ \bibnamefont{Voter}},\ }%
  \bibfield{title}{%
  \enquote{\bibinfo {title} {{Smart Darting Monte Carlo}},}\ }%
  \bibfield{journal}{%
  \bibinfo {journal} {The Journal of Chemical Physics}\ }%
  \textbf{\bibinfo {volume} {114}},\ \bibinfo {pages} {6994--7000} (\bibinfo
  {year} {2001})%
  \bibAnnoteFile{NoStop}{Ioan2001}%
\bibitem{Mochizuki2014}%
  \BibitemOpen
  \bibfield{author}{%
  \bibinfo {author} {\bibfnamefont{K.}~\bibnamefont{Mochizuki}}, \bibinfo
  {author} {\bibfnamefont{C.~S.}\ \bibnamefont{Whittleston}}, \bibinfo {author}
  {\bibfnamefont{S.}~\bibnamefont{Somani}}, \bibinfo {author}
  {\bibfnamefont{H.}~\bibnamefont{Kusumaatmaja}},\ and\ \bibinfo {author}
  {\bibfnamefont{D.~J.}\ \bibnamefont{Wales}},\ }%
  \bibfield{title}{%
  \enquote{\bibinfo {title} {{A conformational factorisation approach for
  estimating the binding free energies of macromolecules}},}\ }%
  \bibfield{journal}{%
  \bibinfo {journal} {Physical Chemistry Chemical Physics}\ }%
  \textbf{\bibinfo {volume} {16}},\ \bibinfo {pages} {2842--2853} (\bibinfo
  {year} {2014})%
  \bibAnnoteFile{NoStop}{Mochizuki2014}%
\bibitem{Olesen2013}%
  \BibitemOpen
  \bibfield{author}{%
  \bibinfo {author} {\bibfnamefont{S.~W.}\ \bibnamefont{Olesen}}, \bibinfo
  {author} {\bibfnamefont{S.~N.}\ \bibnamefont{Fejer}}, \bibinfo {author}
  {\bibfnamefont{D.}~\bibnamefont{Chakrabarti}},\ and\ \bibinfo {author}
  {\bibfnamefont{D.~J.}\ \bibnamefont{Wales}},\ }%
  \bibfield{title}{%
  \enquote{\bibinfo {title} {{A left-handed building block self-assembles into
  right- and left-handed helices}},}\ }%
  \bibfield{journal}{%
  \bibinfo {journal} {RSC Advances}\ }%
  \textbf{\bibinfo {volume} {3}},\ \bibinfo {pages} {12905--12908} (\bibinfo
  {year} {2013})%
  \bibAnnoteFile{NoStop}{Olesen2013}%
\bibitem{Forman2013}%
  \BibitemOpen
  \bibfield{author}{%
  \bibinfo {author} {\bibfnamefont{C.~J.}\ \bibnamefont{Forman}}, \bibinfo
  {author} {\bibfnamefont{S.~N.}\ \bibnamefont{Fejer}}, \bibinfo {author}
  {\bibfnamefont{D.}~\bibnamefont{Chakrabarti}}, \bibinfo {author}
  {\bibfnamefont{P.~D.}\ \bibnamefont{Barker}},\ and\ \bibinfo {author}
  {\bibfnamefont{D.~J.}\ \bibnamefont{Wales}},\ }%
  \bibfield{title}{%
  \enquote{\bibinfo {title} {{Local Frustration Determines Molecular and
  Macroscopic Helix Structures}},}\ }%
  \bibfield{journal}{%
  \bibinfo {journal} {The Journal of Physical Chemistry B}\ }%
  \textbf{\bibinfo {volume} {117}},\ \bibinfo {pages} {7918--7928} (\bibinfo
  {year} {2013})%
  \bibAnnoteFile{NoStop}{Forman2013}%
\bibitem{Bunker2000}%
  \BibitemOpen
  \bibfield{author}{%
  \bibinfo {author} {\bibfnamefont{A.}~\bibnamefont{Bunker}}\ and\ \bibinfo
  {author} {\bibfnamefont{B.}~\bibnamefont{D\"unweg}},\ }%
  \bibfield{title}{%
  \enquote{\bibinfo {title} {{Parallel excluded volume tempering for polymer
  melts}},}\ }%
  \bibfield{journal}{%
  \bibinfo {journal} {Physical Review E}\ }%
  \textbf{\bibinfo {volume} {63}},\ \bibinfo {pages} {016701} (\bibinfo {year}
  {2000})%
  \bibAnnoteFile{NoStop}{Bunker2000}%
\bibitem{Fukunishi2002}%
  \BibitemOpen
  \bibfield{author}{%
  \bibinfo {author} {\bibfnamefont{H.}~\bibnamefont{Fukunishi}}, \bibinfo
  {author} {\bibfnamefont{O.}~\bibnamefont{Watanabe}},\ and\ \bibinfo {author}
  {\bibfnamefont{S.}~\bibnamefont{Takada}},\ }%
  \bibfield{title}{%
  \enquote{\bibinfo {title} {{On the Hamiltonian replica exchange method for
  efficient sampling of biomolecular systems: Application to protein structure
  prediction}},}\ }%
  \bibfield{journal}{%
  \bibinfo {journal} {The Journal of Chemical Physics}\ }%
  \textbf{\bibinfo {volume} {116}},\ \bibinfo {pages} {9058} (\bibinfo {year}
  {2002})%
  \bibAnnoteFile{NoStop}{Fukunishi2002}%
\bibitem{Martiniani2013}%
  \BibitemOpen
  \bibfield{author}{%
  \bibinfo {author} {\bibfnamefont{S.}~\bibnamefont{Martiniani}}\ and\ \bibinfo
  {author} {\bibfnamefont{J.D.}\ \bibnamefont{Stevenson}},\ }%
  \bibfield{journal}{%
  \bibinfo {journal} {Nested Sampling}}%
   (\bibinfo {year} {2013}),\ \url{https://github.com/js850/nested_sampling}%
  \bibAnnoteFile{NoStop}{Martiniani2013}%
\bibitem{Martiniani2013a}%
  \BibitemOpen
  \bibfield{author}{%
  \bibinfo {author} {\bibfnamefont{S.}~\bibnamefont{Martiniani}}\ and\ \bibinfo
  {author} {\bibfnamefont{J.D.}\ \bibnamefont{Stevenson}},\ }%
  \bibfield{journal}{%
  \bibinfo {journal} {Superposition Enhanced Nested Sampling}}%
   (\bibinfo {year} {2013}),\ \url{https://github.com/smcantab/sens}%
  \bibAnnoteFile{NoStop}{Martiniani2013a}%
\end{thebibliography}%

\renewcommand\appendixpagename{Supplementary Information}
\renewcommand\appendixname{S.I.}
\appendix
\appendixpage
\label{sec:SI}

\section{Sampling configurations in a harmonic well}

Given a harmonic potential the configurational density of states for a basin
can be obtained by inverse Laplace transforming the configurational partition function. 
In particular, the scaling goes as:
\begin{equation}
\label{eq:hsa_dos}
g_c(E) = \mathcal{L}^{-1} \left \{ Z_c(\beta) \right \} \propto (E_c - V)^{\frac{\kappa}{2}-1},
\end{equation}
where all the terms that do not depend on energy have been left out. $g_c(E)$ is the configurational density of states, $Z_c$ the configurational partition function (evidence), $\beta = 1/k_BT$ is the inverse temperature, $E_c$ the configurational energy, $V$ the potential energy of the corresponding minimum and $\kappa$ is the number of degrees of freedom (for a $n$-atoms cluster $\kappa=3n-6$). We can write the energy of the simple harmonic oscillator as
\begin{equation}
\label{hooks}
E_c = V + \frac{1}{2} \xi r^2
\end{equation}
where $r$ is the magnitude of the displacement vector and $\xi$ is the stiffness of 
the harmonic spring.
We want to determine the probability distribution of the configurational energy
as a function of the displacement vector norm, $\xi^{\frac{1}{2}} r$, to perform analytical
uniform sampling in the harmonic well. The unnormalised probability of finding
a configuration between $E_c$ and $E_c + \dif E_c$ must be proportional to the
configurational density of states, from Eq.~(\ref{eq:hsa_dos}):
\begin{equation}
  p(E_c) \dif E_c \propto (E_c - V)^{\frac{\kappa}{2}-1} \dif E_c.
\end{equation}
Denoting $q = \xi^{\frac{1}{2}} r$, by a simple change of variables we can
express the energy probability distribution in terms of $q$:
\begin{equation}
p(E_c)\dif E_c = p(E_c(q)) J \dif q \propto  (q^2)^{\frac{\kappa}{2}-1} q \dif q,
\end{equation}
where the Jacobian $J = \dif E_c/\dif q = q$ and the equality simplifies
to the probability density function
\begin{equation}
p(q) \dif q \propto q^{\kappa-1} \dif q .
\end{equation}
Hence $q$ must be distributed according to the power law cumulative distribution function
$P(q) = q^{\kappa}$ (denoted $\text{Pow}(\kappa)$) to obtain the correct
distribution of energies.

In order to sample uniformly below some energy constraint $E_{max}$ we first
generate a random $\kappa -$dimensional vector $\vect{v}$ with norm $v \sim
\text{Pow}(\kappa) \in (0,1]$ in the unit hypersphere. Then starting from
Eq.~(\ref{hooks}) we write
\begin{equation}
\vect{q}_{usc} = \sqrt{\frac{2(E_{max} - V)}{\xi}} \vect{v}
\label{eq:usc}
\end{equation}
where $\vect{q}_{usc}$ is the uniformly sampled configuration vector with
energy $E_c$. It can easily be verified that $\vect{q}_{usc}$ has the correct
inner product:
\begin{equation}
E_c = \frac{1}{2} \xi q_{usc}^2 = (E_{max}-V) v^2,
\end{equation}
where again $v \sim \text{Pow}(\kappa) \in (0,1]$. The configuration $\vect{q}_{usc}$ is then projected onto the orthonormal eigenvector basis $\{\vect{e}_{i}\}$ of the minimum (obtained by diagonalization of the Hessian matrix). The analytically sampled configuration is then:
\begin{equation}
  \vect{r} = \vect{r}_{min} + \sum_{i=1}^{\kappa} \vect{q}_i \vect{e}_i
\end{equation}
where $\vect{r}_{min}$ is the configuration of the minimum stored in the database and $\vect{q}_i$ is the uniformly sampled configuration in Eq.~(\ref{eq:usc}) with $\xi$ being the eigenvalue associated with the $i$-th eigenvector, $\vect{e}_i$.

\section{Hamiltonian replica exchange moves satisfy detailed balance}

The Hamiltonian replica exchange method associates different Hamiltonians (energy functions) with different replicas of the same system, rather than different temperatures. The detailed balance condition for this method has been derived in Ref.~\cite{Bunker2000,Fukunishi2002}.   

The probability to accept an exchange between a configuration $\vect{R}_{sys}$, sampled according to an arbitrary system Hamiltonian $\mathbb{H}_{sys}$, and a configuration $\vect{R}_{har}$ sampled according to a harmonic Hamiltonian $\mathbb{H}_{har}$ is:
\begin{equation}
P\left(\vect{R}_{sys} \rightarrow \vect{R}_{har} \right) = \text{min}\left(1,e^{-\beta(E_{\text{after}}-E_{\text{before}})}\right),
\end{equation}
where
\begin{align}
E_{\text{before}} &= \mathbb{H}_{sys}(\vect{R}_{sys}) + \mathbb{H}_{har}(\vect{R}_{har}),\\
E_{\text{after}} &= \mathbb{H}_{sys}(\vect{R}_{har}) + \mathbb{H}_{har}(\vect{R}_{sys}). 
\end{align}
In SENS we effectively sample configurations at infinite temperature ($\beta \rightarrow 0$), with a hard constraint in energy at $E_{max}$ which can be cast into the Hamiltonian so that 
\[ \mathbb{H}_{x}'(\vect{R}_y) = \left\{ 
  \begin{array}{l l}
    \mathbb{H}_{x}(\vect{R}_{y}) & \quad \text{if}~ \mathbb{H}_{x}(\vect{R}_{y}) \leq E_{max}\\
    \infty \equiv E_{\text{barrier}} & \quad \text{if}~ \mathbb{H}_{x}(\vect{R}_{y})>E_{max}
  \end{array} \right.\]
and we require that
\begin{equation}
1/\beta \gg \mathbb{H}_{sys}(\vect{R}_{sys}),\mathbb{H}_{har}(\vect{R}_{har}).
\end{equation}     
Therefore, if either $\mathbb{H}_{sys}(\vect{R}_{har})$ or $\mathbb{H}_{har}(\vect{R}_{sys})$ diverges, $E_{\text{after}}$ will diverge such that $\beta E_{\text{after}}=\infty$. This last condition requires that
\begin{equation}
E_{\text{barrier}} \gg 1/\beta.
\end{equation}     
It follows that Eq.~(\ref{eq:detailed_balance}) is the correct condition for a swap and that detailed balance holds.  

\section{Onset function scaling}

\begin{figure*}[t]
\setcounter{equation}{1}
\begin{equation}
\label{eq:pdf_sampleb}
p_b(i) = \frac{\Omega_c^{(b)}}{\Omega_c^{(a)}+\Omega_c^{(b)}} = 
\frac{(E_i - V_b)^{\frac{\kappa}{2}} \Theta(E_i-V_b)\prod_{\alpha=1}^{\kappa}\nu_{\alpha}^{(a)} o_a}
{(E_i - V_b)^{\frac{\kappa}{2}}\Theta(E_i-V_b)o_a\prod_{\alpha=1}^{\kappa}\nu_{\alpha}^{(a)} + (E_i -
V_a)^{\frac{\kappa}{2}}\Theta(E_i-V_a)o_b\prod_{\alpha=1}^{\kappa}\nu_{\alpha}^{(b)} }.
\end{equation}
\end{figure*}
\setcounter{equation}{0}

How well would SENS perform in a multifunneled landscape? Let us assume that
SENS is using a database that stores the lowest minima $m_a$ and $m_b$ of two
different funnels, with associated energies $V_a$ and $V_b$, respectively.
Assuming that funnel $F_b$ has already been missed ($F_b$ is not populated and $E_{max}$ is lower than the lowest transition state that leads to $F_{b}$) and all replicas are already
in funnel $F_a$, the probability of successfully sampling in $F_b$ is then
\begin{equation}
\text{Pr}(\text{success}|K~\text{replicas in}~F_a) = 1-\prod_{i=1}^n(1-P_{DS}p_{b}(i))^\mathcal{P},
\end{equation}
\setcounter{equation}{2}
where $n$ is the number of iterations before the calculation terminates,
$\mathcal{P}$ is the number of processors, $P_{DS}$ is a user defined
probability of sampling from the database, and $p_{b}(i)$ is the discrete
probability density that the minimum sampled from the database is $m_b$.
Assuming that the funnels are harmonic we obtain Eq.~(\ref{eq:pdf_sampleb})
where $\nu_{\alpha}=\omega_{\alpha}/(2\pi)$ is the vibrational frequency of mode $\alpha$ and for an object corresponding to a point group with $o$ independent symmetry operations, there are $o$ permutation-inversion operations associated with barrierless reorientations~\cite{Wales2003}. The ideal value for $P_{DS}$ must then satisfy an identity of the form

\begin{figure}[!t]
\centering
\begin{subfigure}
\centering
\includegraphics[width=\linewidth]{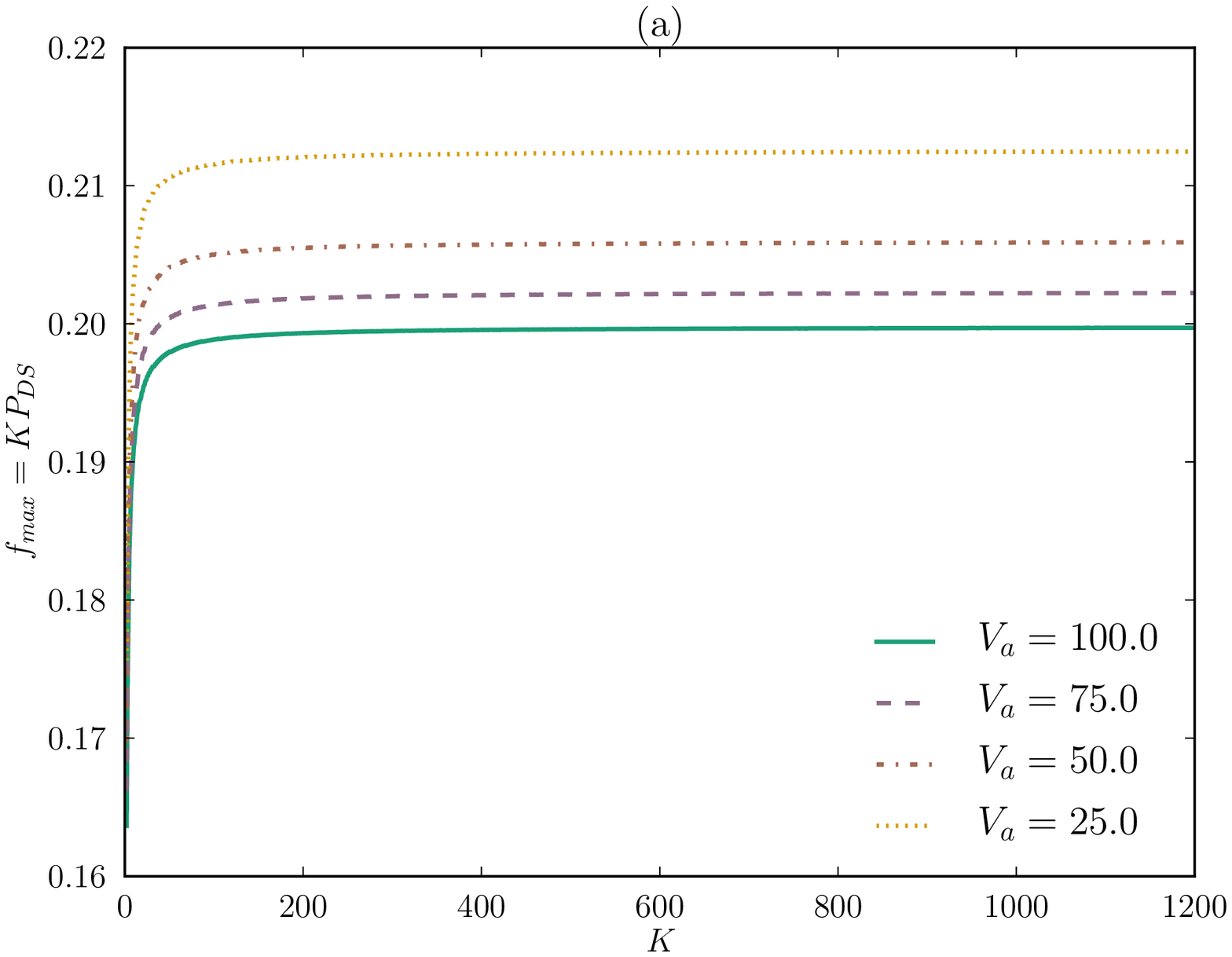}
\end{subfigure}
\begin{subfigure}
\centering
\includegraphics[width=\linewidth]{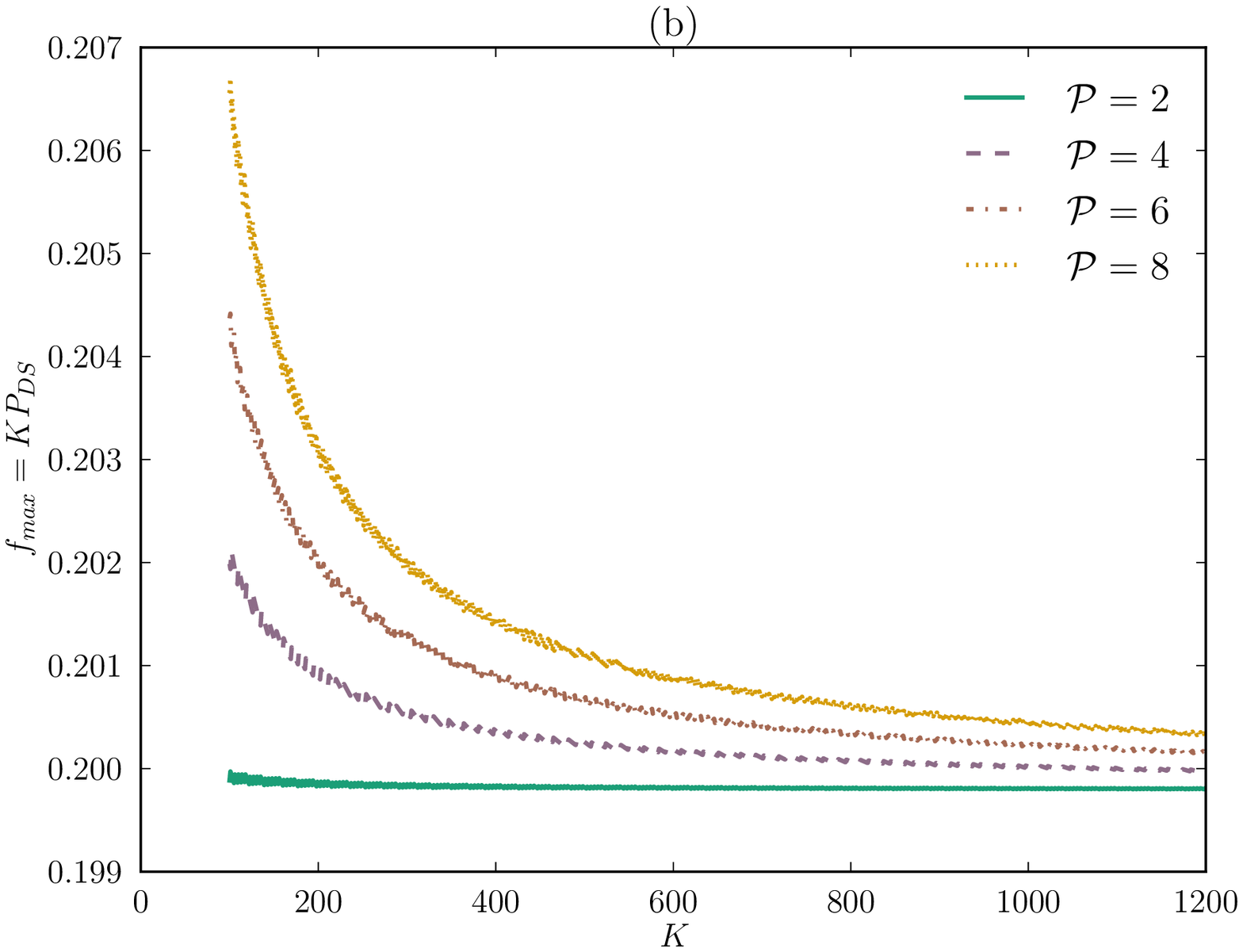}
\caption{Eq.~(\ref{eq:opt_n_steps}) gives the average number of steps necessary
to descend a particular harmonic basin as a function of the number of replicas
$K$ and by substituting it in Eq.~(\ref{eq:ideal_pds}) we can evaluate
$P_{DS}$, the optimal probability of sampling from the database of minima. As
long as $K$ is sufficiently large, $f_{max} = K P_{DS}$ will be approximately
constant, hence the optimal $P_{DS}$ scales as $1/K$. (a) As $V_a$, the
potential of $m_a$, decreases, the volume of $F_a$ increases, and hence the optimal
$f_{max}$ increases as well. (b) The optimal $f_{max}$ should be independent of the number of processors used, hence for sufficiently large $K$ all curves approach the same
value. Unless specified assume $\mathcal{P}=1$, $\kappa = 3$, $E_g = 500$, $V_a
= 100$, $V_b = 0$, $v_{\alpha}^{(a)}=1$, $v_{\alpha}^{(b)}=10$, $o_b = o_a = 1$,
$\delta=10$, $\phi=0.999$.}
\end{subfigure}
\label{fig:ideal_a_P}
\end{figure}

\begin{equation}
\label{eq:ideal_pds}
  1-\prod_{i=1}^{n(V_a)}(1-P_{DS}p_{b}(i))^\mathcal{P} = \phi,
\end{equation}
where $n(V_a)$ is the average number of steps necessary for descending from
$E_g$ to $V_a$ and $\phi$ is a probability close to $1$, say $\phi=0.999$.
Taking the logarithm of both sides, Eq.~(\ref{eq:ideal_pds}) can be rewritten
as
\begin{equation}
\sum_{i=1}^{n(V_a)}\log(1-P_{DS}p_b(i)) = \frac{\log(1-\phi)}{\mathcal{P}}.
\end{equation}
For small $P_{DS}$ linearisation then leads to
\begin{equation}
P_{DS} = -\frac{\log(1-\phi)}{\mathcal{P}\sum_{i=1}^{n(V_a)}p_b(i)},
\end{equation}
which provides an optimal value for $P_{DS}$. To calculate the average number
of steps necessary to descend a harmonic basin, first we calculate an
expression for the set of energies that would be obtained by nested sampling if
at each step the configurational space was compressed exactly by $\mu$. We
note that
\begin{equation}
\mu = \frac{\Omega(E_{i+1})}{\Omega(E_i)}=\frac{(E_{i+1}-V)^{\frac{\kappa}{2}}}{(E_{i}-V)^{\frac{\kappa}{2}}}
\end{equation}
from which we find
\begin{equation}
\label{eq:energy_analitical}
E_i = (E_0-V)\mu^{\frac{2i}{\kappa}}+V,
\end{equation}
where $\mu = 1 - \mathcal{P}/(K+1)$. The number of steps $n(V_a)$
necessary to descend from $E_g$ to $E-V_a=10^{-\delta}$ can be obtained by
rearranging Eq.~(\ref{eq:energy_analitical}) to give
\begin{equation}
\label{eq:opt_n_steps}
n(V_a) = -\frac{\kappa}{2}\frac{\delta \log(10)+\log(E_g-V_a)}{\log(\mu)}.
\end{equation}
Finally, substituting Eq.~(\ref{eq:energy_analitical}) for $E_i$ in
Eq.~(\ref{eq:pdf_sampleb}) we can evaluate Eq.~(\ref{eq:ideal_pds}) numerically
to obtain an optimal value for $P_{DS}$ (approximating $n(V_a)$ to the nearest
integer).

In the main text we introduce the onset function Eq.~(\ref{eq:onset_function})
and suggest that for small $P_{DS}$ an optimal way to 
make the probability of sampling from the
database of minima independent of the number of replicas, is to use the
prefactor $1/K$, hence the maximum frequency to sample from the minima, should
be $f_{max}/K$, where $f_{max}$ is a user-defined parameter.
Eq.~(\ref{eq:opt_n_steps}) gives the average number of steps necessary to
descend a particular harmonic basin as a function of the number of replicas
and by substituting this result in Eq.~(\ref{eq:ideal_pds}) we can evaluate $P_{DS}$,
the optimal probability of sampling from the database of minima. In
Fig.~\ref{fig:ideal_a_P} we plot $f_{max} = K P_{DS}$ vs $K$. For large $K$ the
optimal value of sampling from the minima scales as $1/K$, thus justifying the
use of the prefactor in the onset function.

\section{Algorithms}
\subsection{SENS}
A complete pseudo-code implementation of the SENS algorithm is provided in Algorithm~\ref{alg:sens}.

\begin{algorithm}
\begin{algorithmic}[1]
\LineComment{initialisation, set $i=0$}
\State generate a database of minima \In \db ;
\For{ minimum \In \db}
\State compute the Hessian matrix and its eigenvalues
\Statex[1] (needed to compute the HSA weight);
\EndFor
\While{$i<K$}
\State sample a random configuration of the system;
\State store its coordinates and its energy in \LList ;
\EndWhile
\LineComment{main loop}
\While{termination condition \Is \False}

\State remove the $\mathcal{P}$ replicas $\{ {\bf R}_{m}^{(1)},\dots , {\bf R}_{m}^{(\mathcal{P})} \} \equiv \{ {\bf R}_{m} \}$ with  
\Statex[1] highest energy $\{ E_{m}^{(1)} > \dots > E_{m}^{(\mathcal{P})} \} \equiv \{ E_{m} \}$ 
\Statex[1] from \LList ;

\State append $\{ E_{m} \}$ to \OList ;
\State set $E_{max} = E_{m}^{(\mathcal{P})}$ ;

\State sample $\mathcal{P}$ replicas $\{ {\bf R}_{s}^{(1)},\dots,{\bf R}_{s}^{(\mathcal{P})} \} \equiv \{{\bf R}_s\} $ from 
\Statex[1]\LList at random;

\State add a copy of $\{{\bf R}_s\}$ to \LList ;
\State \sens{$\{\{{\bf R}_s\}$,~\db,~$E_{max}\}$} ;
\EndWhile
\State append \LList to \OList ;
\end{algorithmic}
\caption{Superposition Enhanced Nested Sampling}
\label{alg:sens}
\end{algorithm}

\subsection{Exact SENS}

A pseudo-code implementation of the
\texttt{MCLoop} function for exact SENS can be found in
Algorithm~\ref{alg:exact_sens_mcloop}.

\begin{algorithm}
\begin{algorithmic}[1]
\ForAll{${\bf R}_s$ \In $\{{\bf R}_s\}$}
\State sample a minimum $m$ from \db according to 
\Statex[1] its HSA entropic weight;

\State analytically generate a configuration ${\bf R}_{har}$ in 
\Statex[1] the harmonic well of $m$, with energy 
\Statex[1] $\mathbb{H}_{har}({\bf R}_{har})=E_{har}^{({\bf R}_{har})}$;

\State evaluate $\mathbb{H}_{sys}({\bf R}_{har})=E_{sys}^{({\bf R}_{har})}$
\Statex[1] and $\mathbb{H}_{har}({\bf R}_s)=E_{har}^{({\bf R}_s)}$\;
\If{$E_{har}^{({\bf R}_s)} \leq E_{max}$ {\bf and} $E_{sys}^{({\bf R}_{har})} \leq E_{max}$}
\State \emph{swap} ${\bf R}_{har} \leftrightarrow {\bf R}_s$;
\For{$l=0$ \To $N$-steps}
\State walk ${\bf R}_s$ by sampling uniformly within 
\Statex[3] $\{ E_{s} \} \leq E_{max}$;
\EndFor
\Else{ reject the swap and perform a standard MCMC;}
\EndIf
\EndFor
\end{algorithmic}
\caption{Exact SENS \texttt{MCLoop}}
\label{alg:exact_sens_mcloop}
\end{algorithm}

\noindent The energy of a configuration ${\bf R}_s$ with respect to ${\bf R}_{min}$ is
\begin{equation}
E_{har}^{({\bf R}_s)} = E_{har}^{({\bf R}_{min})} + ({\bf R}_s-{\bf R}_{min})^{\top}{\bf \mathcal{H}}_{min}({\bf R}_s-{\bf R}_{min}),
\end{equation}
where ${\bf \mathcal{H}}_{min}$ is the Hessian matrix associated with ${\bf R}_{min}$. Note
that more careful considerations are needed when dealing with Lennard-Jones
clusters, in fact ${\bf R}_{min}$ is fixed in the database (hence also ${\bf \mathcal{H}}_{min}$), thus
breaking the translational and rotational invariance of the system. In order to
avoid this problem we can either align ${\bf R}_{min}$ to ${\bf R}_s$, but then
${\bf \mathcal{H}}_{min}$ must also be recalculated, or we can do the opposite, which
is the most efficient solution. A pseudo-code implementation of the \texttt{MCLoop} 
function specific to this system is provided in Algorithm~\ref{alg:exact_sens_cluster}.

\begin{algorithm}[]

\begin{algorithmic}[1]
\LineComment{this loop is performed in parallel}
\ForAll{${\bf R}_s$ \In $\{{\bf R}_s\}$}


\State sample a minimum $m$ from \db according to 
\Statex[1] its HSA entropic weight;

\State analytically generate a configuration ${\bf R}_{har}$ in 
\Statex[1] the harmonic well of $m$, with energy 
\Statex[1] $\mathbb{H}_{har}({\bf R}_{har})=E_{har}^{({\bf R}_{har})}$;

\State evaluate $\mathbb{H}_{LJ}({\bf R}_{har}) = E_{LJ}^{({\bf R}_{har})}$;

\If{$E_{LJ}^{({\bf R}_{har})} > E_{max}$}
\State reject the exchange and perform a standard 
\Statex[2] MCMC;
\Else

\State quench ${\bf R}_s \rightarrow {\bf R}_{que}$ with energy 
\Statex[2] $\mathbb{H}_{LJ}({\bf R}_{que})=E_{LJ}^{({\bf R}_{que})}$ and find the minimum 
\Statex[2] configuration ${\bf R}_{min}$ with the corresponding energy
\Statex[2] in \db;

\State find the set of permutations ($\mathcal{P}$) and rotations ($\mathcal{R}$) 
\Statex[2] such that $\mathcal{PR}{\bf R}_{que} = {\bf R}_{min}$;

\State generate ${\bf R}^{(\mathcal{PR})}_s=\mathcal{PR}{\bf R}_s$, which is now aligned 
\Statex[2] to ${\bf R}_{min}$;
\State compute $\mathbb{H}_{har}({\bf R}_s^{(\mathcal{PR})}) = E_{har}^{({\bf R}_s^{(\mathcal{PR})})}$;
\If{$E_{har}^{({\bf R}^{(\mathcal{PR})}_s)} > E_{max}$}
\State reject the move and perform a standard 
\Statex[3] MCMC;
\Else{ \emph{swap} ${\bf R}_{har} \leftrightarrow {\bf R}_s$;}
\For{$l=0$ \To $N$-steps}
\State walk ${\bf R}_s$ by sampling uniformly within 
\Statex[4] $\{ E_{s} \} \leq E_{max}$;
\EndFor
\EndIf
\EndIf
\EndFor
\end{algorithmic}

\caption{Exact SENS \texttt{MCLoop} for LJ-clusters}
\label{alg:exact_sens_cluster}
\end{algorithm}

\subsection{Approximate SENS}
A pseudo-code implementation of the
\texttt{MCLoop} function for approximate SENS is presented in
Algorithm~\ref{alg:approximate_sens_mcloop}.

\begin{algorithm}
\begin{algorithmic}
\ForAll{${\bf R}_s$ \In $\{{\bf R}_s\}$}
\State sample a minimum configuration ${\bf R}_{b}$ from \db 
\Statex[1] according to its HSA entropic weight;

\If{$u \sim \text{Uniform}(0,1) < P_{DS}$}
\State \emph{swap} ${\bf R}_{b} \leftrightarrow {\bf R}_s$;
\For{$l=0$ \To $N$-steps}
\State walk ${\bf R}_s$ by sampling uniformly within 
\Statex[3] $\{ E_{s} \} \leq E_{max}$;
\EndFor
\Else{ reject the swap and perform a standard MCMC;}
\EndIf
\EndFor
\end{algorithmic}
\caption{Approximate \texttt{MCLoop}}
\label{alg:approximate_sens_mcloop}
\end{algorithm}

\newpage

\section{Statistical Uncertainty by Compression Factor Resampling}
\label{sec:err_compr_fact}
The nested sampling algorithm produces as its primary product a list of parameters (in our case energies) with an associated fraction of configurational space $X_i=\prod_{j=0}^it_j$, where the $t_j$ are the compression factors sampled on a unit interval with probability distribution $\text{Beta}(K-\mathcal{P}+1,\mathcal{P})$ and expectation value $\mu = 1 - \mathcal{P}/(K+1)$. The exploration of configuration space is the challenging and time consuming part of the algorithm, while the overhead due to the assignment of compression factors is almost irrelevant. In general we use the expectation value $\mu$ of these compression factors in order to find the bins of density of states, $g$, that we need to calculate thermodynamic properties. Given a set of energies obtained by nested sampling, the correct size of the bins for the density of states is one unique realisation of the compression factors $t$ that we do not know a priori (it is for this reason that we use the expectation value $\mu$). There is some statistical uncertainty associated with the distribution of the bin size, which is ultimately due to the distribution of compression factors $t$, which we know. Since we are interested in the distribution of some observable $Q(\vect{E})$, say the heat capacity, we can use a representative set of parameters $\vect{\widetilde{E}} = \widetilde{E}_1, \widetilde{E}_2, \dots, \widetilde{E}_n$ (energies) obtained by nested sampling (the time consuming part) and sample $c$ sets of compression factors $\vect{t}_{\ell}=t_1^{(\ell)},t_2^{(\ell)},\dots,t_n^{(\ell)}$ to associate with this representative set of parameters. This procedure is justified by the fact that we are interested in the probability distribution of $Q(\vect{\widetilde{E}})$ given the joint probability distribution $p(\vect{t}_{\ell}) = p(t_1^{(\ell)})p(t_2^{(\ell)})\dots p(t_n^{(\ell)})$. The mean and variance of $Q(\vect{\widetilde{E}})$ are therefore

\begin{equation}
\left \langle Q(\vect{\widetilde{E}}) \right \rangle = \sum_{\ell=1}^c Q(\vect{\widetilde{E}}) p(\vect{t}_\ell)=\frac{1}{c} \sum_{\ell=1}^c Q(\vect{\widetilde{E}}|\vect{t}_\ell);
\label{eq:err_alpha_expectation}
\end{equation}

\begin{align}
Var\left(Q(\vect{\widetilde{E}})\right) &= \sum_{\ell=1}^c Q(\vect{\widetilde{E}})^2 p(\vect{t}_\ell) - \left(\sum_{\ell=1}^c Q(\vect{\widetilde{E}})^2 p(\vect{t}_\ell)\right)^2 \nonumber \\
\label{eq:err_alpha_variance}
&= \frac{1}{c} \sum_{\ell=1}^c Q(\vect{\widetilde{E}}|\vect{t}_\ell)^2 - \left( \frac{1}{c} \sum_{\ell=1}^c Q(\vect{\widetilde{E}}|\vect{t}_\ell) \right)^2.
\end{align}

A protocol for quantifying uncertainty of this form was already suggested by Skilling~\cite{Skilling2006}. However, this method suffers from some flaws: it does not include an estimate of systematic uncertainties and does not include in any way an estimate of the uncertainty due to the incomplete sampling of configurational space. The latter effect is due to the fact that resampling is performed over one representative set of energies $(\vect{\widetilde{E}})$ that could be missing a whole part of configurational space.

\section{Results}
\subsection{LJ$_{\bf 38}$}

 LJ$_{38}$ has a double-funnel energy landscape~\cite{Doye1995a} and its heat capacity exhibits three peaks: the first high temperature peak corresponds to a vapour-liquid transition, the second peak corresponds to melting, and the low temperature peak corresponds to a solid-solid transition from the cuboctahedral global minimum to the lowest icosahedral minimum~\cite{Doye1998}. The PT heat capacity curve for LJ$_{38}$ generally overestimates the temperature at which the solid-solid phase-like transition occurs and, consequently, disappears under the melting peak, and should appear, instead, as a small shoulder slightly under the melting peak~\cite{Poulain2006,SHARAPOV}. For this reason long calculations are necessary for the heat capacity to converge. Partay et al.~\cite{Partay2012} employed $K=244000$ replicas and O$(10^{12})$ energy evaluations to resolve the heat capacity of LJ$_{38}$ by NS. Fig.~$5$ compares the heat capacity curves obtained by PT (using a box of radius $R=2.8~\sigma$), HSA (computed using $\gtrsim 89000$ minima) and SENS for LJ$_{38}$. SENS was carried out using $K=50000$ replicas and $N=10000$ steps for each MCMC walk. The database of minima used for SENS contained approximately $89000$ minima. It appears that neither version of SENS can outperform NS or parallel tempering, which require the same number of energy evaluations to converge; see Table~III for comparison. Exact SENS fails due to the inaccuracy of the HSA, and Table~V shows the number of effective swaps from and to the basin. These numbers are considerably smaller than for LJ$_{31}$ (see Table~IV), even if the total number of iterations for LJ$_{38}$ is much larger. Approximate SENS should generally work independently of the quality of the HSA. Here, however, the three transitions overlap significantly thus preventing sampling from the database early enough to get the right low temperature behaviour without affecting the high temperature behaviour. This is not the case for LJ$_{75}$ where the phase-like transitions are well separated in temperature and sampling can start early enough to get the correct low temperature behaviour without affecting the melting transition.

\begin{figure}
\centering
\includegraphics[width=\linewidth]{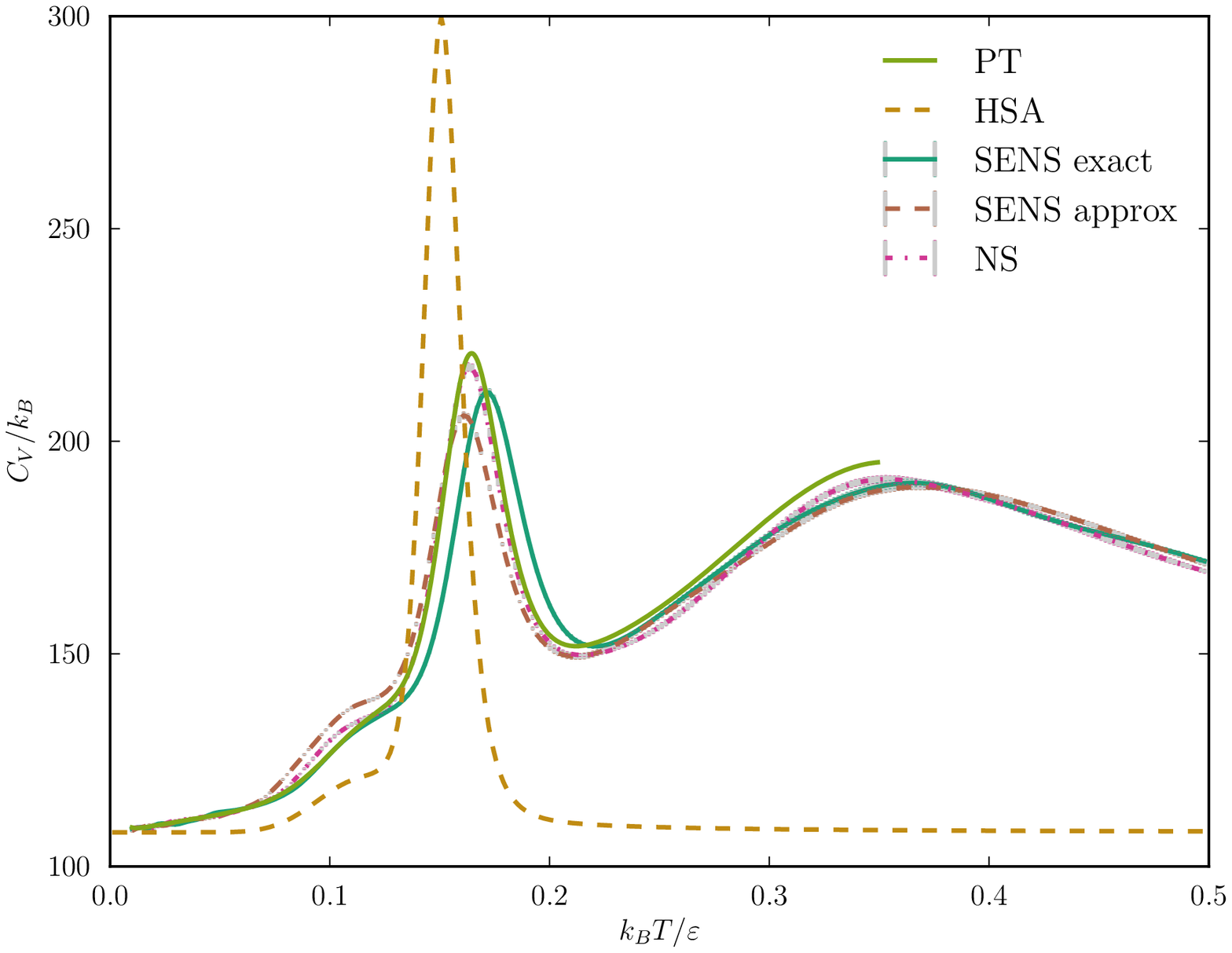}
\caption{Heat capacity curves for LJ$_{38}$. The PT and HSA results were obtained by parallel tempering and the harmonic superposition approximation, respectively.}
\label{fig:cv_lj38}
\end{figure}

\begin{table}
\centering
\begin{tabular}{lccc}
\firsthline
\hline
\multicolumn{4}{c}{LJ$_{38}$} \\
\hline
Method      & $K$         & $N$        & $N_{\text{E}}^{(total)}$ \\
\hline
\hline
PT          &              &               & $O(10^{11})$ \\
NS ref.\cite{Partay2012} & $244000$ &  & $O(10^{12})$ \\ 
NS          & $50000$     & $10000$      & $3.3 \times 10^{11}$               \\
SENS approx & $50000$     & $10000$      & $3.3 \times 10^{11}$ \\
SENS exact  & $50000$     & $10000$      & $3.3 \times 10^{11}$  \\
\lasthline
\end{tabular}
\caption{Comparison of methods used to obtain the LJ$_{31}$ heat capacity curves shown in Fig.~$5$. $N_{\text{E}}^{(total)}$ indicates the total number of energy evaluations (summed over all processors).}
\label{table:lj38}
\end{table}

\subsection{Swap statistics in exact SENS calculations}

In this section we present statistics for the exact SENS swaps in the longest runs of each LJ system presented in the paper. For each minimum we report the total number of swaps that led to this minimum, the number of effective swaps to this minimum (thus excluding the swaps within the same minimum) and the number of effective swaps that led from the minimum to another.

\begin{table*}
\centering
\begin{tabular}{lccc}
\firsthline
\hline
\multicolumn{4}{c}{LJ$_{31}$} \\
\hline
minimum      & total swaps to & effective swaps from          & effective swaps to    \\
\hline
\hline
$-133.586421919$          & $346376$  & $4225$    & $5994$  \\
$-133.293821966$          & $14585$   & $7167$    & $5403$  \\
$-133.183574005$          & $1529$    & $1325$    & $1325$  \\
$-133.104620445$          & $412$     & $385$     & $365$   \\
$-132.998423589$          & $25$      & $11$      & $25$    \\
$-132.801757275$          & $1$       & $0$       & $1$     \\ 
$-132.765536037$          & $0$       & $1$       & $0$     \\ 
$-132.721370719$          & $1$       & $0$       & $1$     \\
\lasthline
\end{tabular}
\label{table:lj31_swaps}
\caption{Number of swaps to and from basins in exact SENS for LJ$_{31}$ using $K=20000$ replicas, $N=10000$ steps for each MCMC and $\mathcal{P}=16$. Total number of iterations per processor$=650025$.}
\end{table*}

\begin{table*}
\centering
\begin{tabular}{lccc}
\firsthline
\hline
\multicolumn{4}{c}{LJ$_{38}$} \\
\hline
minimum      & total swaps to & effective swaps from          & effective swaps to    \\
\hline
\hline
$-173.928426591$ & $486584$ & $245$ & $48$   \\
$-173.252378416$ & $65$ & $26$ & $63$ \\
$-173.134317009$ & $163$ & $27$ & $154$ \\
$-172.958633408$ & $4$ & $2$ & $4$ \\
$-172.877736411$ & $38$ & $6$ & $37$ \\
$-172.234926493$ & $0$ & $1$ & $0$ \\
$-171.992596189$ & $1$ & $0$ & $1$ \\
\lasthline
\end{tabular}
\caption{Number of swaps to and from basins in exact SENS for LJ$_{38}$ using $K=50000$ replicas, $N=10000$ steps for each MCMC and $\mathcal{P}=16$. Total number of iterations per processor$=2082691$.}
\label{table:lj38_swaps}
\end{table*}

\begin{table*}
\centering
\begin{tabular}{lccc}
\firsthline
\hline
\multicolumn{4}{c}{LJ$_{75}$} \\
\hline
minimum      & total swaps to & effective swaps from          & effective swaps to    \\
\hline
\hline
$-397.492330983$ & $632019$  & $1$  & $704$ \\
$-396.282248826$ & $15$ & $296$ & $12$      \\
$-396.238512215$ & $14$ & $270$ & $10$      \\
$-396.193034959$ & $7$ & $41$ & $5$         \\
$-396.192994186$ & $3$ & $11$ & $3$         \\
$-396.191648856$ & $4$ & $45$ & $4$         \\
$-396.186860193$ & $0$ & $1$ & $0$          \\
$-396.126268882$ & $2$ & $63$ & $1$         \\
$-396.061061075$ & $1$ & $0$ & $1$          \\
$-396.061598578$ & $0$ & $13$ & $0$         \\
$-396.061061075$ & $0$ & $1$ & $0$          \\
$-396.057293139$ & $1$ & $0$ & $1$          \\
$-395.992783183$ & $1$ & $0$ & $1$          \\
\lasthline
\end{tabular}
\caption{Number of swaps to and from basins in exact SENS for LJ$_{75}$ using $K=60000$ replicas, $N=10000$ steps for each MCMC and $\mathcal{P}=16$. Total number of iterations per processor$=5043100$.}
\label{table:lj75_swaps}
\end{table*}

\end{document}